\documentclass{article}
\usepackage{url}

\usepackage{amsmath}
\usepackage{graphicx}
\usepackage{enumerate}
\usepackage{amssymb}
\usepackage{mathtools}
\usepackage{booktabs}
\usepackage{support-caption} 
\usepackage{subcaption, caption}
\usepackage{placeins} 
\usepackage{mathrsfs}
\usepackage{amsfonts}
\usepackage{bm}
\usepackage{amsthm}
\newtheorem{definition}{Definition}
\newtheorem{proposition}{Proposition}
\usepackage [english]{babel}
\usepackage [autostyle, english = american]{csquotes}
\MakeOuterQuote{"}
\DeclareGraphicsExtensions{.png,.pdf}
\usepackage{biblatex}
\def\tblhead#1{\hline\\[-9pt]#1\\\hline\\[-9.75pt]}
\def\lastline{\\\hline}

\begin{document}

\title{An intersectional framework for counterfactual fairness in risk prediction}

\author{SOLVEJG WASTVEDT$^\ast$, JARED D. HULING, JULIAN WOLFSON\\[4pt]
\textit{Division of Biostatistics, University of Minnesota, Minneapolis, MN}
\\[2pt]
{wastv004@umn.edu}}

\markboth%
{S. Wastvedt and others}
{Intersectional counterfactual fairness}

\maketitle

\footnotetext{To whom correspondence should be addressed.}

\begin{abstract}
{Along with the increasing availability of health data has come the rise of data-driven models to inform decision-making and policy. These models have the potential to benefit both patients and health care providers but can also exacerbate health inequities. Existing "algorithmic fairness" methods for measuring and correcting model bias fall short of what is needed for health policy in two key ways. First, methods typically focus on a single grouping along which discrimination may occur rather than considering multiple, intersecting groups. Second, in clinical applications, risk prediction is typically used to guide treatment, creating distinct statistical issues that invalidate most existing techniques. We present summary unfairness metrics that build on existing techniques in "counterfactual fairness" to address both challenges. We also develop a complete framework of estimation and inference tools for our metrics, including the unfairness value ("u-value"), used to determine the relative extremity of unfairness, and standard errors and confidence intervals employing an alternative to the standard bootstrap. We demonstrate application of our framework to a COVID-19 risk prediction model deployed in a major Midwestern health system.}
\end{abstract}

\section{Introduction}
\label{sec:introduction}

Along with the increasing availability of data in many sectors has come the rise of data-driven models to inform decision-making by predicting outcomes like disease, hospital readmission, and many more. In health care, such \emph{risk prediction models} are used to assess patients' likelihood of certain adverse outcomes and guide assignment of treatments or interventions. These models have the potential to benefit both patients and health care providers by personalizing treatment and improving efficiency (\cite{kilicArtificialIntelligenceMachine2020}).

However, risk prediction models also have the potential to entrench or exacerbate health inequities. High-profile examples of inequities perpetrated by the models, which often use opaque, "black box" machine learning techniques, have emerged both in health care contexts and elsewhere (\cite{obermeyerDissectingRacialBias2019a}). In response, techniques for measuring and correcting model bias, which broadly constitute the field of "algorithmic fairness", have proliferated. These techniques include an array of definitions of model unfairness (\cite{castelnovoClarificationNuancesFairness2022}), most of which compare some measure of performance of the model across groups defined by a social characteristic like race or gender or an attribute such as age, income, etc.

For clinical applications, existing algorithmic fairness work falls short in two key ways. First, definitions typically focus on a single grouping along which discrimination may occur, for example assessing performance for men vs. women. This simplification fails to recognize that discrimination occurs based on many different groupings which interact in complex ways. Second, in clinical applications, risk prediction is typically used to guide treatment, and use of a treatment presents distinct statistical issues that invalidate most existing fairness measurement techniques. To our knowledge, no existing fairness work addresses both challenges; in this paper we present three novel unfairness metrics that are valid when the risk score is used to guide treatment and account for intersecting forms of discrimination. We also develop a complete framework for estimation and inference on our metrics. For estimation, we present new procedures that restrict the error rate estimates underlying our metrics within the natural $[0,1]$ bounds. We then present tools for inference on the summary metrics, including the \emph{unfairness value ("u-value")}, used to determine the relative extremity of an unfairness measurement, and standard errors and confidence intervals. Our standard error procedure employs an alternative to the standard bootstrap necessitated by the fact that our metrics are aggregations of absolute values.

\subsection{Intersectionality in health care and risk prediction}

Our work falls in the emerging area of techniques addressing multiple, intersecting forms of discrimination which is typically referred to as \emph{intersectional fairness} work. The term \emph{intersectionality} was coined by legal scholar Kimberl\'e Crenshaw, although its ideas have a long history in Black feminist thought and elsewhere (\cite{nashBlackFeminismReimagined2019, hillcollinsIntersectionality2016}). Crenshaw used intersectionality to describe the experiences of oppression Black women face, arguing that this oppression is not simply additive, or the sum of racism and sexism, but a distinct form of discrimination that demands new analysis (\cite{crenshawDemarginalizingIntersectionRace1989, crenshawMappingMarginsIntersectionality1991}). Definitions of what exactly constitutes intersectionality are often debated, but it has been broadly characterized as a "knowledge project" engaged in by scholars and social justice activists from a wide array of disciplines whose focus is on "power relations and social inequalities" (\cite{hillcollinsIntersectionalityDefinitionalDilemmas2015}). Intersectional analysis seeks to identify and dismantle interlocking forms of discrimination wherever they occur, whether based on race, sex, gender identity, class, or otherwise (\cite{intersectionalitytraininginstituteWhatIntersectionality2022}).
    
In health care, the importance of an intersectional approach has been firmly established by work documenting the health impacts of interlocking forms of discrimination. As a recent example, the COVID-19 pandemic has disproportionately hurt groups at the intersections of racial, gender and economic oppression in many ways, including disproportionate health consequences for Black Americans (\cite{aleemNewCDCData2020}) and job losses unequally borne by Latina women (\cite{alexiafernandezcampbellEvenPositiveJobs2020}). Clinical risk models are susceptible to all these interlocking inequities. For example, researchers at a major health system decided against implementing a patient "no-show" prediction tool after realizing that it could discriminate both explicitly and implicitly based on a wide range of patient characteristics (\cite{murrayDiscriminationArtificialIntelligence2020}).

\subsection{The need for counterfactual fairness}

In clinical settings, risk prediction typically informs and is accompanied by assignment of some treatment, for example care coordination services or intensive COVID-19 therapies. In this setting, we wish to predict the risk to a patient were they to remain untreated and thereby target our intervention to those at highest risk. However, when a treatment is in use, data available for fairness assessment consists of a mix of outcomes observed with treatment and without. Moreover, if the treatment is more often applied or is more effective in certain groups, the mismatch between desired prediction targets and available data is more severe for these groups. Conventional methods for assessing performance will fail in this situation, especially for groups where the mismatch is greatest. Fairness measurements that fail to account for this phenomenon will also fail and could falsely ascribe fairness to a model that in fact discriminates against one group. 

\subsubsection{Using potential outcomes to measure performance}

This failure is the motivation for a recent, non-intersectional approach to algorithmic fairness called \emph{counterfactual fairness} which uses techniques from causal inference (\cite{costonCounterfactualRiskAssessments2020, mishlerFairnessRiskAssessment2021a}). These metrics replace observed outcomes with \emph{potential outcomes}, or the outcomes that would have been observed under a given treatment decision. Using potential outcomes allows us to assess a model's performance and fairness relative to a baseline in which no patients are treated, thus avoiding the bias described above.

\subsubsection{Related work}
This counterfactual framework differs from a similarly named area of work in which counterfactuals are defined with respect to the protected characteristic rather than treatment assignment (e.g. \cite{kusnerCounterfactualFairness2017}), and which considers potential outcomes in a world in which a person belonged to a different group, e.g. were of a different race or gender. However, the validity of causal assumptions in this approach is difficult to establish, particularly in the case of socially constructed characteristics like race or gender (\cite{fawkesSelectionIgnorabilityChallenges2022, rubinStatisticsCausalInference1986}). Further, as \cite{costonCounterfactualRiskAssessments2020} discuss, this is not the relevant counterfactual for settings in which risk predictions guide interventions, so we do not take this approach.

Elsewhere, existing work has proposed measurements for fairness across multiple intersecting characteristics (e.g. \cite{fouldsIntersectionalDefinitionFairness2020}) and methods for training fairer models (e.g.  \cite{hebert-johnsonMulticalibrationCalibrationComputationallyIdentifiable2018}), but these methods do not account for the bias due to treatment described above.

\section{Existing definitions and statistical framework} \label{sec:related-work}

In this section, we describe a statistical framework for measuring counterfactual equalized odds with a single protected characteristic (\cite{mishlerFairnessRiskAssessment2021a}) and our extension of this framework to the intersectional context.

Following convention in algorithmic fairness work, define a \emph{protected characteristic} as any grouping, such as race or gender, along which we wish to measure discrimination. Let $A$ denote a protected characteristic, which \cite{mishlerFairnessRiskAssessment2021a} assume is binary. Let $S$ denote a binary risk prediction, which may be obtained from a model that produces predicted probabilities through selection of a cutpoint. If a range of possible cutpoints are of interest, the user can apply our methods on the series and evaluate the fairness at each threshold. We leave cutpoint selection to the user since the appropriate threshold will differ depending on the application. Let $D$ denote a binary treatment assignment and $Y$ a binary outcome such as an adverse health event. Under a binary treatment, there are two potential outcomes: $Y^0$, the outcome under no treatment, and $Y^1$, the outcome under treatment. The $Y^0$ outcome is the most relevant quantity in many clinical risk prediction settings, where we aim to predict patients' baseline risk in order to guide treatment. The following counterfactual quantities, defined in \cite{mishlerFairnessRiskAssessment2021a} and \cite{costonCounterfactualRiskAssessments2020}, substitute the $Y^0$ potential outcome for $Y$ in common model performance and fairness metrics.

\begin{definition} \label{def:cferr}
The \emph{counterfactual false positive rate} of a prediction $S$ for protected group $A=a$, denoted $cFPR(S,a)$, is equal to $Pr(S=1 | Y^0 = 0, A = a)$. The \emph{counterfactual false negative rate}, $cFNR(S,a)$, is equal to $Pr(S=0|Y^0 = 1, A = a)$. 
\end{definition}

These quantities are counterfactual analogues to the commonly used false positive and false negative rates, referred to as $\emph{observational error rates}$. 

When $D$, $S$, $A$, and $Y^0$ are binary, equality of the counterfactual false positive rate and counterfactual false negative rate between the levels of $A$ is equivalent to a counterfactual version of the common fairness metric \emph{equalized odds}.

\begin{definition} \label{def:ceqodds}
For $A,D,S,Y^0 \in \{0,1\}$, a prediction $S$ satisfies \emph{counterfactual equalized odds} if $cFPR(S,0) = cFPR(S,1)$ and $cFNR(S,0) = cFNR(S,1)$.
\end{definition}

We take counterfactual equalized odds as the starting point for our metrics because of its relevance to clinical settings, where a typical performance metric of interest is a model's ability to correctly identify patients who need treatment (minimize counterfactual false negatives) and not erroneously recommend treatment for patients who do not need it (minimize counterfactual false positives). A prediction model satisfying counterfactual equalized odds is "fair" in the sense that its ability to perform both of these tasks is the same across protected characteristic groups.

To quantify counterfactual fairness, \cite{mishlerFairnessRiskAssessment2021a} define the \emph{counterfactual error rate differences} of a prediction $S$, denoted $\Delta^+(S)$ and $\Delta^-(S)$, as the differences in counterfactual false positive and false negative rates between two groups of the single protected characteristic $A \in \{0,1\}$. In symbols, $\Delta^+(S) = cFPR(S,0) - cFPR(S,1)$ and $\Delta^-(S) = cFNR(S,0) - cFNR(S,1)$

\section{Quantifying intersectional counterfactual unfairness} \label{sec:definitions}

We now extend counterfactual error rates to the intersectional context and propose three novel summary metrics that can be used to assess the unfairness of a risk prediction model. In all estimands, we accommodate the intersectional context by replacing the single protected characteristic, $A$, with the vector $\bm{A}$. Let $m$ be the number of protected characteristics that we wish to consider. Denote the characteristics $A_j$, $j \in \{1, ..., m\}$ and assume each $A_j$ is a categorical variable with a finite number of levels, the set of which is denoted $\mathcal{A}_j$. Let $\bm{A} = \{A_1, A_2, ..., A_m\}^T \in \mathcal{A}$ contain all protected characteristics of interest, where $\mathcal{A}$ is the set of all possible combinations of all levels of the $m$ characteristics.

Under this notation, we denote the counterfactual error rate differences between the group having protected characteristic vector $\bm{a}$ and the group having $\bm{a}'$ as $\Delta^+(S, \bm{a}, \bm{a}')$ (difference in counterfactual false positive rates) and $\Delta^-(S, \bm{a}, \bm{a'})$ (difference in counterfactual false negative rates). These are defined as $\Delta^+(S, \bm{a}, \bm{a}') = cFPR(S,\bm{a}) - cFPR(S,\bm{a}')$ and $\Delta^-(S, \bm{a}, \bm{a}') =  cFNR(S,\bm{a}) - cFNR(S,\bm{a}')$.

We use the general $\Delta(S, \bm{a}, \bm{a}')$ throughout to denote either the positive ($\Delta^+(S, \bm{a}, \bm{a}')$) or negative ($\Delta^-(S, \bm{a}, \bm{a'})$) version of the counterfactual error rate difference. The relative consequences of false positives and false negatives can vary widely by situation, so the version of the metric that is of most interest will also vary. The general notation $\Delta(S, \bm{a}, \bm{a}')$ also opens the possibility that a metric other than $cFPR$ or $cFNR$ could be compared, such as area under a counterfactual version of the receiver operating characteristic (ROC) curve, although we leave this extension to future work.

\subsection{Proposed summary unfairness metrics} \label{sec:sub-listdefs}

In this section, we propose metrics for aggregating counterfactual unfairness across many intersecting protected groups. Our metrics use simple summaries; however, for concreteness we define them here as they apply to the counterfactual error rate differences. First, we give an overall picture of unfairness among all groups by taking the average. 

\begin{definition}
\emph{Average intersectional counterfactual unfairness} is the average of absolute error rate differences across all possible pairs of protected characteristic vectors.

\begin{equation} \label{def_avg}
    \Delta_{AVG}(S) = \frac{1}{(||\mathcal{A}||-1)||\mathcal{A}||/2} \sum_{\bm{a}, \bm{a}' \in \mathcal{A}} \left | \Delta (S,\bm{a}, \bm{a}') \right | 
\end{equation}

where $||\mathcal{A}||$ denotes the cardinality of the set $\mathcal{A}$.
\end{definition}

When one group displays extreme unfairness and differences among all other groups are small, $\Delta_{AVG}$ is pulled downwards by the smaller differences and fails to capture the full extent of the unfairness. The problem is worse if the number of intersectional groups is large, since then $\Delta_{AVG}$ has even less sensitivity to unfairness involving only one or a few groups. To provide a fuller picture, we use the maximum to highlight the most extreme error rate difference.

\begin{definition}
\emph{Maximum intersectional counterfactual unfairness} is the maximum absolute error rate difference across all combinations of protected characteristics.

\begin{equation} \label{def_max}
    \Delta_{MAX}(S) = \max_{\bm{a}, \bm{a}' \in \mathcal{A}} \left | \Delta(S,\bm{a}, \bm{a}') \right |
\end{equation}
\end{definition}

In some contexts, it may be desirable to capture changes in the spread of error rate differences rather than simply their relative sizes. If error rates are equally spaced, such that there are relatively large differences among all groups, this suggests a different issue than if one extreme difference is driving the measurement. Our third metric takes the variance to highlight these issues.

\begin{definition}
\emph{Variational intersectional counterfactual unfairness} is the variance of absolute error rate differences across all combinations of protected characteristics.

\begin{equation} \label{def_sd}
    \Delta_{VAR}(S) = \frac{1}{(||\mathcal{A}||-1)||\mathcal{A}||/2 - 1} \sum_{\bm{a}, \bm{a}' \in \mathcal{A}} \left( \left | \Delta (S,\bm{a}, \bm{a}') \right | - \Delta_{AVG} \right)^2 
\end{equation}
\end{definition}

\subsection{Advantage of intersectional metrics} \label{sec:sub-demonstration}
Here we demonstrate the need for our approach by comparing it to an alternative that follows immediately from existing work and is counterfactual but not intersectional. This "marginal" metric adds additional protected characteristics, in a non-intersecting manner, to the counterfactual error rate differences metric proposed by \cite{mishlerFairnessRiskAssessment2021a}. Let $\mathcal{A}^*$ be the count of pairs within protected groups, i.e. $\mathcal{A}^* = \sum_{j=1}^m {\binom{||\mathcal{A}_j||}{2}}$. Then $\Delta_{MARG}(S) = \frac{1}{\mathcal{A}^*} \sum_{j=1}^m \sum_{a_j, a_j' \in \mathcal{A}_j} \left | \Delta (S,a_j, a_j') \right |$.

We consider a simplified data generating scenario with two binary protected characteristics, one level of each substantially less common, creating one majority group, one minority, and two moderately sized groups. We assume a model with poorest performance, as measured by highest counterfactual false negative rate, for the minority group, better performance for the moderately sized groups, and best for the majority group.

\begin{figure}[h]
    \centering
    \includegraphics[width=.9\textwidth]{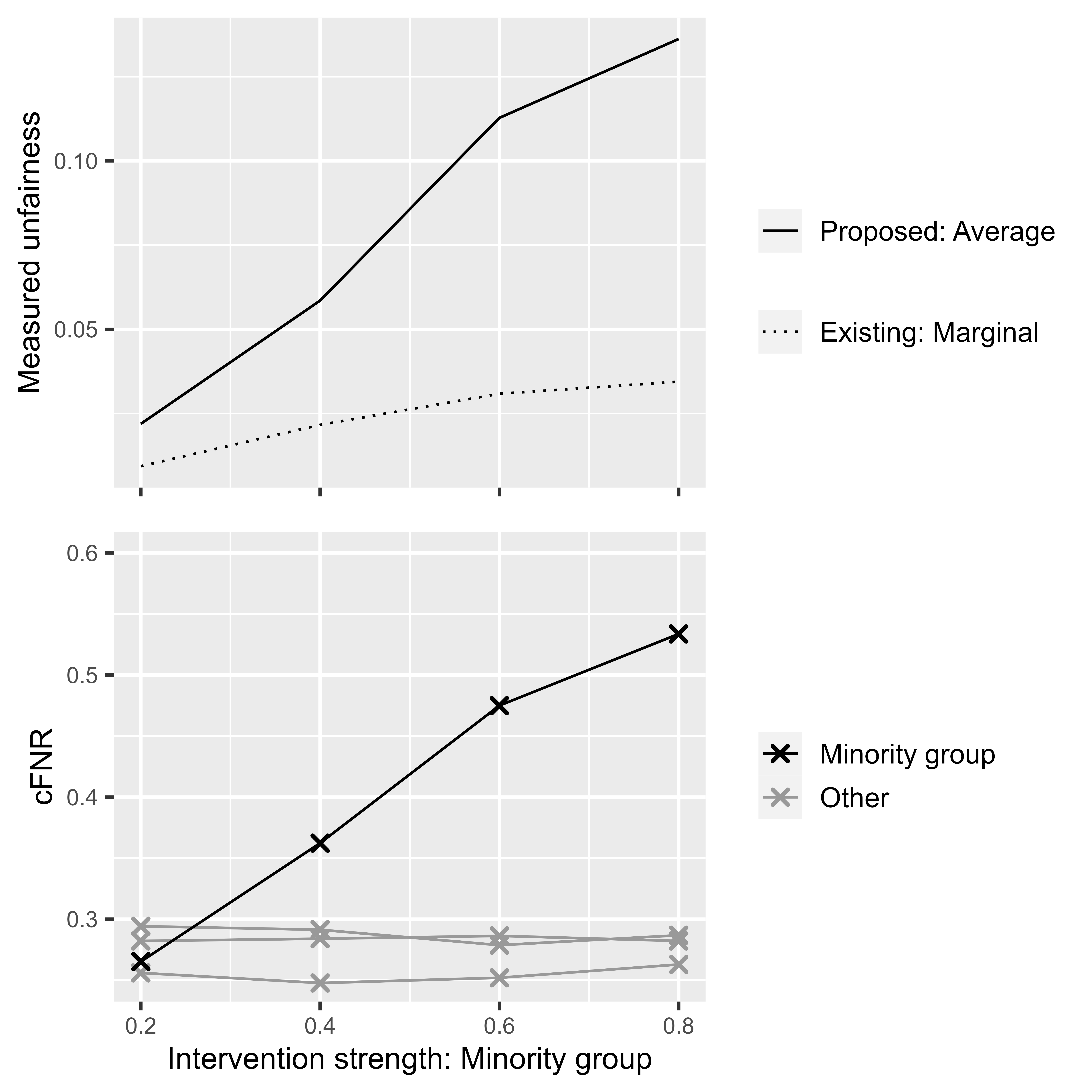}
    \caption{Proposed $\Delta_{AVG}$ vs. $\Delta_{MARG}$. Top panel: Values of metrics using the true group-specific error rates of the risk model as established with a validation set of size $50,000$. Bottom panel: Group-specific true $cFNR$ values as established using the validation set.}
    \label{fig:simtrue_b}
\end{figure}

Figure \ref{fig:simtrue_b} shows scenarios with increasing counterfactual false negative rate ($cFNR$) for the minority group, moving right along the x-axis, while the $cFNR$ for the other groups does not change (Figure \ref{fig:simtrue_b}, bottom panel). Details of data generation for each of these scenarios are given in the Supplementary Material. As shown, $\Delta_{MARG}$ (dotted line) fails to fully capture the increasing unfairness involving this small group. In contrast, our proposed $\Delta_{AVG}$ (solid line) provides a clearer picture of the increasing unfairness and its magnitude. To understand this effect, note that $\Delta_{AVG}$ weights unfairness from all groups equally, while $\Delta_{MARG}$ down-weights unfairness in the smallest groups. In our example, this translates to $\Delta_{MARG}$ under representing the increasing levels of unfairness in the smallest group.

\FloatBarrier

\section{Identification} \label{sec:identification}

Existing work by \cite{mishlerFairnessRiskAssessment2021a} has shown that the counterfactual error rates of a binary prediction $S$ can be identified for a single binary protected characteristic, $A$, under the standard causal inference assumptions of consistency, positivity, and ignorability. We make these same assumptions, substituting the vector $\bm{A}$ for the single protected characteristic $A$.

 However, the doubly robust estimator \cite{mishlerFairnessRiskAssessment2021a} propose for the positive counterfactual error rate can exceed the bounds $[0,1]$, constraints an error rate estimate would ideally respect. We present a slightly different identification result that leads to an estimator which respects these bounds; our result follows the derivation in \cite{mishlerFairnessRiskAssessment2021a} for $cFNR(S,\pmb{a})$ but differs for $cFPR(S,\pmb{a})$. Proposition \ref{prop:identification} shows the general result for a prediction $S$ and counterfactual outcome $Y^0$. A proof of Proposition \ref{prop:identification}, including our assumptions using the protected characteristic vector notation, is given in the Supplementary Material.

\begin{proposition} \label{prop:identification}
Given arbitrary functions with finite mean $f(S)$ and $g(Y^0)$, $E[f(S)g(Y^0)I(\bm{A} = \bm{a})] = E\left[ \frac{(1-D)f(S)g(Y)I(\bm{A}=\bm{a})}{1-\pi(\bm{A},\bm{X},S)} \right ]$.
\end{proposition}

We then apply Proposition \ref{prop:identification} to identification of the counterfactual error rates of the prediction $S$ for the group having protected characteristic vector $\bm{A} = \bm{a}$. Let $p[f(S),g(Y)] = E\left[ \frac{(1-D)f(S)g(Y)I(\bm{A}=\bm{a})}{1-\pi(\bm{A},\bm{X},S)} \right ]$. Then $cFPR(S, \bm{a}) = p[S,1-Y]/p[1,1-Y]$ and $cFNR(S, \bm{a}) = p[1-S,Y]/p[1,Y]$.

The error rate differences, $\Delta(S, \bm{a}, \bm{a}')$, are then identified as $\Delta^+(S, \bm{a}, \bm{a}') = \big |cFPR(S,\bm{a}) - cFPR(S,\bm{a}')\big |$ and $\Delta^-(S, \bm{a}, \bm{a}') = \big |cFNR(S,\bm{a}) - cFNR(S,\bm{a}')\big |$. Our proposed unfairness metrics are identified by inserting the relevant error rate differences into equations \eqref{def_avg}, \eqref{def_max}, and \eqref{def_sd}.

\section{Estimation} \label{sec:estimation}

\subsection{Estimating intersectional, counterfactual unfairness metrics}
To estimate our new counterfactual, intersectional unfairness metrics, we first propose estimators of the counterfactual error rates. Following the notation in Section \ref{sec:related-work}, let $\{\bm{A}_i, D_i, Y_i, \bm{X}_i, S_i\}$, $i = 1, ..., n$ be the observed data and binary predictions from the risk model.

\begin{align}
    \widehat{cFPR}(S,\bm{a}) &= \frac{\sum_{i=1}^n[(1-D_i)S_i(1-Y_i) I\{\bm{A}_i=\bm{a}\}/(1-\hat{\pi}_i)]}{\sum_{i=1}^n[(1-D_i) (1-Y_i)I\{\bm{A}_i=\bm{a}\}/(1-\hat{\pi}_i)]} \label{cfpr_est} \\
    & \nonumber \\
    \widehat{cFNR}(S,\bm{a}) &= \frac{\sum_{i=1}^n[(1-D_i)(1-S_i)Y_iI\{\bm{A}_i=\bm{a}\}/(1-\hat{\pi}_i)]}{\sum_{i=1}^n[(1-D_i)Y_iI\{\bm{A}_i=\bm{a}\}/(1-\hat{\pi}_i)]} \label{cfnr_est}
\end{align}

To estimate the error rate differences, we replace $cFPR(S,\bm{a})$ and $cFNR(S,\bm{a})$ in $\Delta^+(S, \bm{a}, \bm{a}')$ and $\Delta^-(S, \bm{a}, \bm{a}')$ with their estimates. Finally, to estimate our new counterfactual, intersectional unfairness metrics, we use the estimated error rate differences in equations \eqref{def_avg}, \eqref{def_max}, and \eqref{def_sd}.

\begin{align}
    \widehat{\Delta}_{AVG}(S) &= \frac{1}{(||\mathcal{A}||-1)||\mathcal{A}||} \sum_{\bm{a}, \bm{a}' \in \mathcal{A}} \left | \widehat{\Delta} (S,\bm{a}, \bm{a}')\right | \label{est_avg} \\
    \widehat{\Delta}_{MAX}(S) &= \max_{\bm{a}, \bm{a}' \in \mathcal{A}} \left | \widehat{\Delta}(S,\bm{a}, \bm{a}') \right | \label{est_max} \\
    \widehat{\Delta}_{VAR}(S) &= \frac{1}{||\mathcal{A}|| - 1} \sum_{\bm{a}, \bm{a}' \in \mathcal{A}} \left( \left | \widehat{\Delta} (S,\bm{a}, \bm{a}') \right | - \widehat{\Delta}_{AVG} \right)^2  \label{est_sd}
\end{align}

We estimate the non-intersectional metric $\Delta_{MARG}$ and the non-counterfactual metric $\Delta_{OBS}$ for comparison in an analogous manner.

\subsection{Estimating nuisance parameters} \label{sec:estimation-nuissance}
We suggest two options for estimation of the propensity score function $\pi(\bm{A}, \bm{X}, S)$: a simple generalized linear model and a more flexible ensemble approach that combines multiple machine learning-based estimates with a parametric model, such as the Super Learner (\cite{laanSuperLearner2007}). As with all IPW estimators, our estimators assume a correctly specified propensity score model that is $\sqrt{n}$ convergent and lies in a Donsker class \cite{vandervaartWeakConvergenceEmpirical1996}. If a machine learning propensity score model is used, one could use such a model combined with a parametric model in an ensemble estimated via sample splitting techniques (\cite{naimiChallengesObtainingValid2020}). We use a 10-fold cross-fitting procedure for our ensemble model that contains both a machine learning estimator and a parametric model, in which the model is fitted on the remaining data to obtain predictions for the held-out fold. While in practice, use of sample splitting and an ensemble that includes a parametric model improves performance, we caution that the use of a flexible nonparametric approach like this does not meet theoretical conditions needed by singly-robust IPW estimators.

It is well-established that in contrast to weighted estimators, doubly robust estimators, which include models of both the propensity score and outcome, need only correctly specify one of these models in order to achieve consistency and have less stringent requirements to obtain inferential guarantees. As such, in the Supplementary Material, we present doubly robust versions of our estimators that are constrained in $[0,1]$ using the techniques of \cite{robinsCommentPerformanceDoubleRobust2007}. However, we maintain weighted estimators as our primary method for two reasons. First, our estimators require practitioners to specify one model rather than two; because of this, the estimators can also easily be applied to multiple prediction/outcome combinations without returning to the modeling step. More importantly, in many clinical contexts, practitioners have much more information about the true propensity score model since treatment decisions are based on well-known combinations of variables. In contrast, outcomes, which are typically complex biological processes, are usually much harder to model well. If the outcome model is incorrect, doubly robust estimators similarly rely on the correct specification of the propensity score model as do our weighted estimators.

For comparison, we include two other methods of nuisance parameter estimation in the analyses that follow. In Section \ref{sec:sim}, we use a version of expressions \eqref{cfpr_est} and \eqref{cfnr_est} with the propensity score calculated using the true data generating mechanism as specified in our simulation. We also include a regression estimator, in which the weighted estimates of $Y^0$ (the quantity $(1-Y_i)(1-D_i)/(1-\hat{\pi}_i)$ in equation \eqref{cfpr_est} and the quantity $Y_i(1-D_i)/(1-\hat{\pi}_i)$ in equation \eqref{cfnr_est}) are replaced with regression estimates (derivation in Supplementary Material).

\section{Inference for new unfairness measures} \label{sec:inference}

In this section we propose procedures for assessing the extent of unfairness according to our new metrics. The \emph{unfairness value} that we propose is a high-level comparison between a measurement on one of our new metrics and a hypothetical, perfectly fair model. We propose approaches for estimation of standard errors and confidence intervals to provide further insight into the extent of unfairness.

\subsection{An unfairness value} \label{sec:sub-inference_null}

To assess the relative unfairness of a model, we propose an \emph{unfairness value ("u-value")}, which assesses a model against a reference distribution representing a hypothetical, perfectly fair model. To construct the reference distribution, we jointly permute the protected characteristic vectors, $\bm{A}_i$, $i = 1, ..., n$, of all individuals in the observed data while holding all other data constant. The permutation simulates a situation in which protected characteristic vectors are randomly assigned and thus have no relation to model error rates, i.e. a situation of no unfairness. Since, even with a relatively small sample, the total number of possible permutations of the protected characteristic vectors will be large, we use Monte Carlo approximation and sample with replacement from the set of all possible permutations. We then obtain reference distributions for the metrics in Section \ref{sec:definitions} by calculating each metric on each of the permuted datasets.

In many applications, users will be willing to trade some unfairness for gains in performance and thus will wish to test against a chosen threshold rather than an expectation of zero unfairness. Let $\delta$ be the user-defined acceptable limit for unfairness on a given metric. Then define the \emph{u-value}, $u(S,\delta)$, as the proportion of permutations in which the observed value of the metric exceeds the permuted value by more than $\delta$. 

For example, let $\{\widehat{\Delta}_{AVG}(S)^*_1, ..., \widehat{\Delta}_{AVG}(S)^*_P\}$ be the set of values of $\Delta_{AVG}$ calculated across $P$ permuted datasets. Then 

\begin{equation*}
    u_{AVG}(S,\delta) = \frac{\sum_{j=1}^P I\big(\widehat{\Delta}_{AVG}(S) - \widehat{\Delta}_{AVG}(S)_j^* > \delta\big )}{P}
\end{equation*}

A lower u-value indicates less unfairness, with a statistically significant u-value rejecting the null hypothesis of unfairness exceeding the threshold.
This test is mathematically equivalent to a non-inferiority test (\cite{walkerNoninferiorityStatisticsEquivalence2019}). We have chosen the name "u-value" to emphasize our departure from a sharp null hypothesis of zero unfairness and aid in interpretation, since in this test a higher "unfairness value" means greater unfairness. Constructing the test in this manner may also help avoid "ethics washing" (\cite{weinbergRethinkingFairnessInterdisciplinary2022}) potentially problematic algorithms, since here the null hypothesis is that a model does exceed the unfairness threshold. In contrast, p-values obtained with a traditional null hypothesis of no unfairness may encourage the misinterpretation that a model is fair if the test fails to reject the null.

\subsection{Standard errors}

Because our metrics aggregate absolute values of error rate differences, the null value for each metric is zero, which is on the boundary of the parameter space. It has been established that the standard bootstrap is inconsistent when the true value of the parameter is on the boundary (\cite{andrewsInconsistencyBootstrapWhen2000}). Thus, for any model that is close to fair, using the standard bootstrap with our methods gives incorrect results. Instead, we propose use of the \emph{rescaled bootstrap} (\cite{andrewsInconsistencyBootstrapWhen2000}), a technique generalized as the \emph{numerical delta method} by \cite{hongNumericalDeltaMethod2018}. 

Let $\hat{\theta}_i^*$ be the rescaled bootstrap estimates, such that $\hat{\theta}_i^* = \sqrt{m}(\hat{\theta}_{m,i}^* - \hat{\theta}_n)$, $i = 1, ..., B$ where $B$ is the number of bootstrap resamples and $m$ is the size of the resamples, with $m \rightarrow \infty$ and $m(lnln(n))/n \rightarrow 0$ as $n \rightarrow \infty$. Let $\bar{\theta}_m^*$ be the sample mean of the $\hat{\theta}_i^*$. Then we estimate the standard error of $\hat{\theta}_n$ as follows. First, take the sample variance of the rescaled bootstrap estimates. This sample variance consistently estimates the variance of $\sqrt{n}(\hat{\theta}_n - \theta)$. Then multiply by $1/n$ to obtain an estimate of the variance of $\hat{\theta}_n$. The rescaled bootstrap estimate of $SE(\hat{\theta}_n)$ is therefore $\widehat{SE(\hat{\theta}_n)} = \sqrt{\frac{1}{n}\left [ \frac{\sum_{i=1}^m (\hat{\theta}^*_i - \bar{\theta}_m^*)^2}{m-1} \right ]}$.

\subsection{Confidence intervals} \label{sec:sub-inference_ci}

Using the rescaled bootstrap, we have several options for construction of confidence intervals. We propose an adaptation of the \emph{bootstrap t-interval} adjusted for the rescaling of the bootstrap estimates. In the standard bootstrap case, the bootstrap t-interval uses the distribution of $t^* = (\hat{\theta}^* - \hat{\theta})/\hat{\sigma}^*$ to approximate the distribution of $t = (\hat{\theta} - \theta)/\hat{\sigma}$, where $\hat{\sigma}$ is an estimate of the standard deviation of $\hat{\theta}$ (\cite{boosEssentialStatisticalInference2013}). For the rescaled case, we obtain the values $t^*_m = (\hat{\theta}_{m}^* - \hat{\theta}_n)/\widehat{SE(\hat{\theta}_n)}$, using our estimate of the standard error of $\hat{\theta}$. Denote the empirical distribution function of the $t^*_m$ as $\hat{F}_{t^*}$. Then a rescaled $1-\alpha$ bootstrap t-interval for $\hat{\theta}_n$ is given by $\left \{ \hat{\theta}_n - \widehat{SE(\hat{\theta}_n)} \hat{F}^{-1}_{t^*}(1-\alpha/2), \hat{\theta}_n - \widehat{SE(\hat{\theta}_n)} \hat{F}^{-1}_{t^*}(\alpha/2) \right \}$.

We also consider for comparison two alternative methods: the \emph{normal approximation interval} and the \emph{percentile interval}, which uses percentiles of the rescaled bootstrap estimates, transformed back to the scale of the parameter estimate.

\section{Simulations} \label{sec:sim}

We simulate three scenarios with varying types of unfairness to demonstrate properties of our new metrics and estimators. The first scenario simulates low unfairness, with approximately equal counterfactual error rates for all protected groups. The second scenario simulates unfairness involving many groups. The third scenario simulates unfairness in which one protected group has a large error rate difference with all other groups.

\begin{figure}[h]
    \centering
    \begin{subfigure}[b]{0.37\textwidth}
         \centering
         \includegraphics[width=\textwidth]{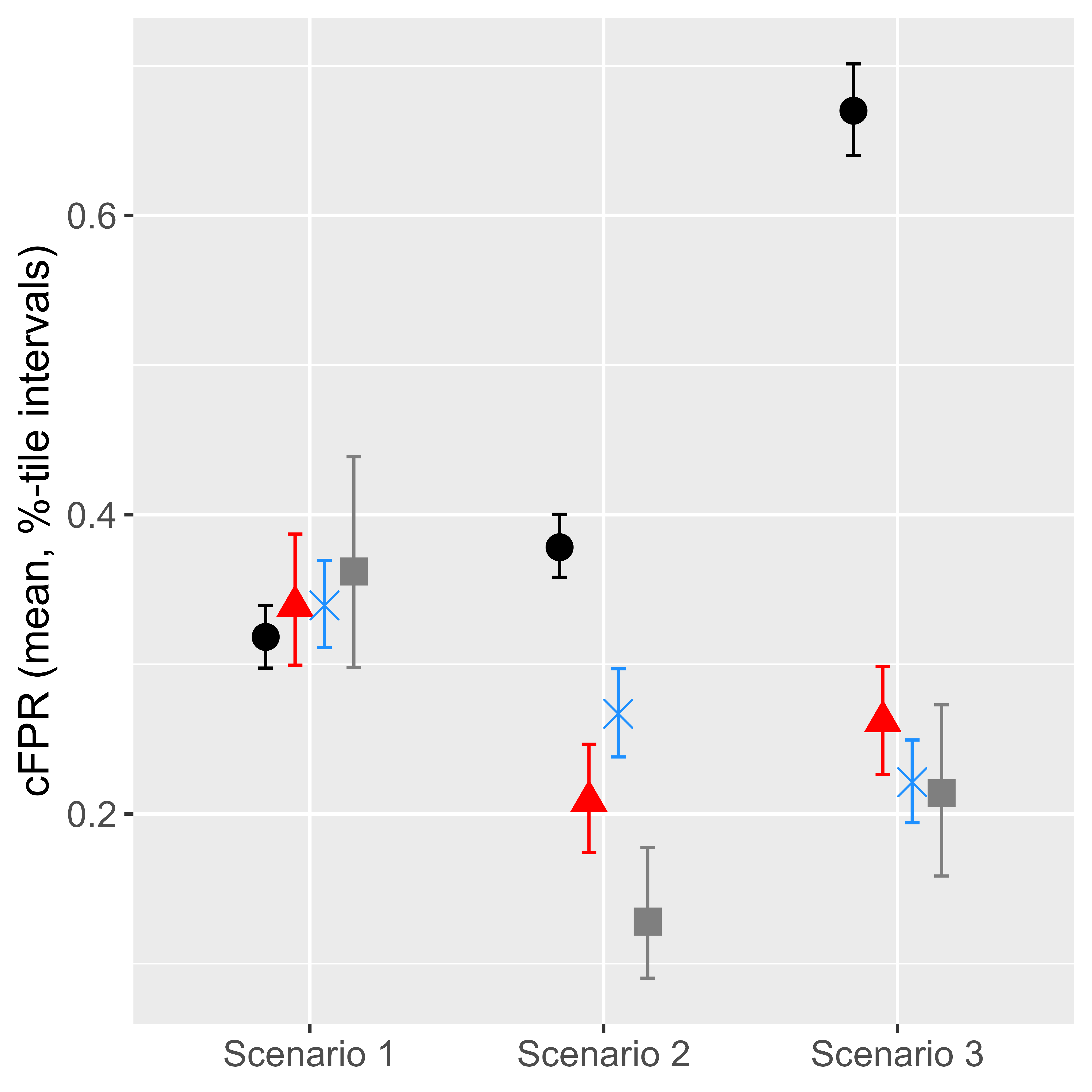}
         \caption{Counterfactual false positive rates}
         \label{fig:sim_errrates_a}
     \end{subfigure}
     \hfill
     \begin{subfigure}[b]{0.5\textwidth}
         \centering
         \includegraphics[width=\textwidth]{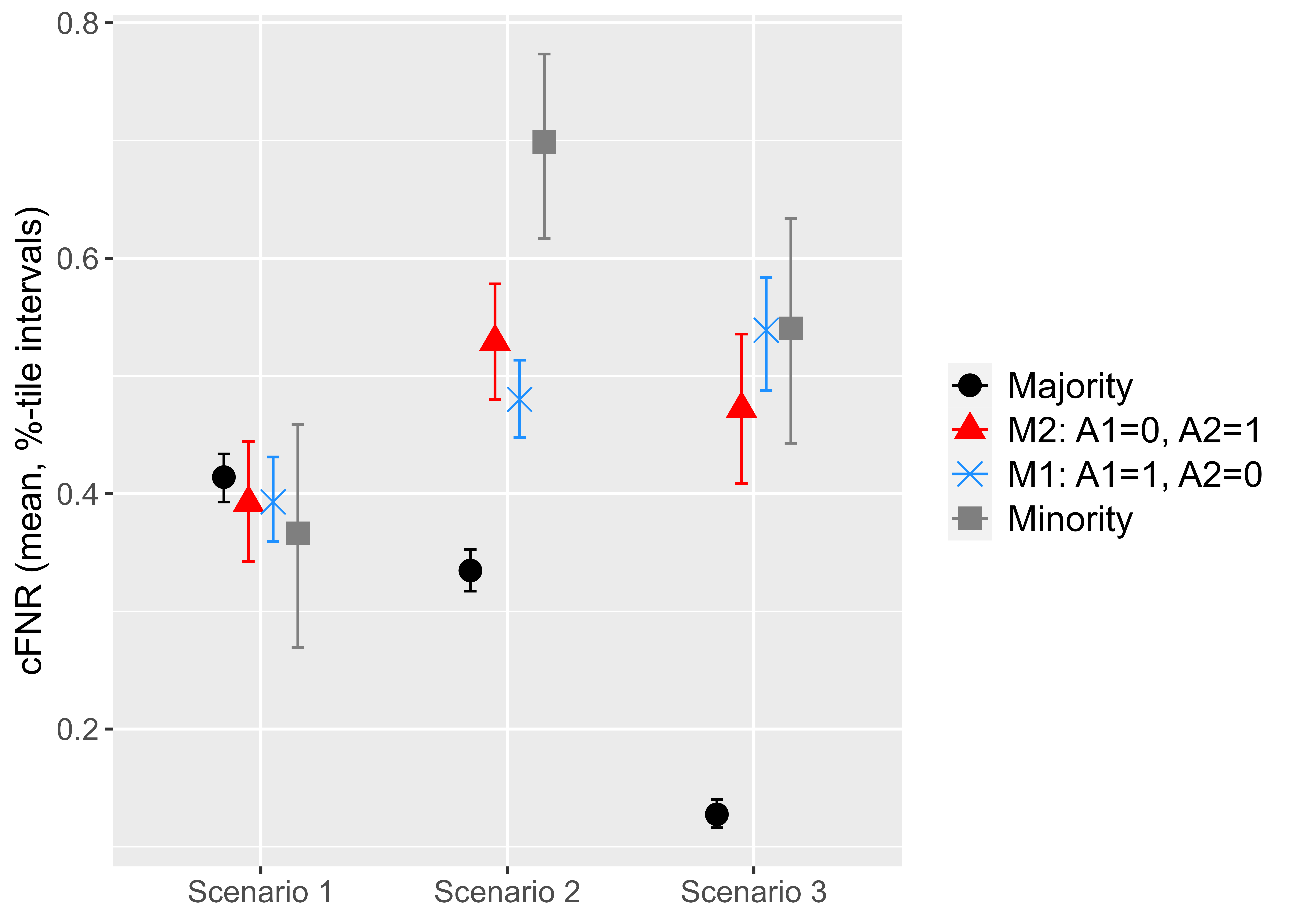}
         \caption{Counterfactual false negative rates}
         \label{fig:sim_errrates_b}
     \end{subfigure}
    \caption{Counterfactual error rates by protected group for each of the unfairness scenarios considered in Section \ref{sec:sim}. Shapes show the mean of $500$ replications of the estimation procedure, with $N_{estimation} = 9,000$ and the GLM propensity score model estimation method. Intervals show the $0.025$ and $0.975$ quantiles of the replications.}
    \label{fig:sim_errrates}
\end{figure}

For all scenarios, we follow the framework in Section \ref{sec:sub-demonstration}, with two binary protected characteristics. We refer to the groups as the "majority", "minority", $M1$ ($A_1 = 1, A_2=0$) and $M2$ ($A_1 = 0, A_2=1$). Figure \ref{fig:sim_errrates} depicts the counterfactual error rates in the three scenarios.

\subsection{Data generation and simulation set-up}
The foundation for our simulations is the data generating procedure described in \cite{mishlerFairnessRiskAssessment2021a}, in which a training data set is generated as the basis for training a risk prediction model. Our training data set is of size $N_{train} = 1,000$, and our risk prediction model is a random forest. We then generate a validation data set with sample size $50,000$ to establish the true error rate properties of the risk model. Finally, we generate what we refer to as the "estimation" data set. We obtain binary risk predictions on this data using the risk prediction model and a classification cutoff of $0.5$. We then perform nuisance parameter estimation and estimate the values of our unfairness metrics. Full details of the data generating process and scenarios are given in the Supplementary Material.

For each of our three unfairness scenarios, we consider four estimation data set sample sizes ($1,000$, $5,000$, $7,000$, $9,000$). We consider the four estimation methods described in Section \ref{sec:estimation}: regression estimation and inverse probability weighting using three propensity score models (GLM, ensemble, true data generating mechanism). For the ensemble propensity score model, we use a Super Learner combining a generalized linear model and a random forest. The regression estimates use a generalized linear model. We perform $500$ replications of the data set generation and estimation procedure for each combination of parameters and unfairness scenario.

\subsection{Comparison of novel unfairness metrics}

\begin{figure}[h]
    \centering
    \begin{subfigure}[b]{0.38\textwidth}
         \centering
         \includegraphics[width=\textwidth]{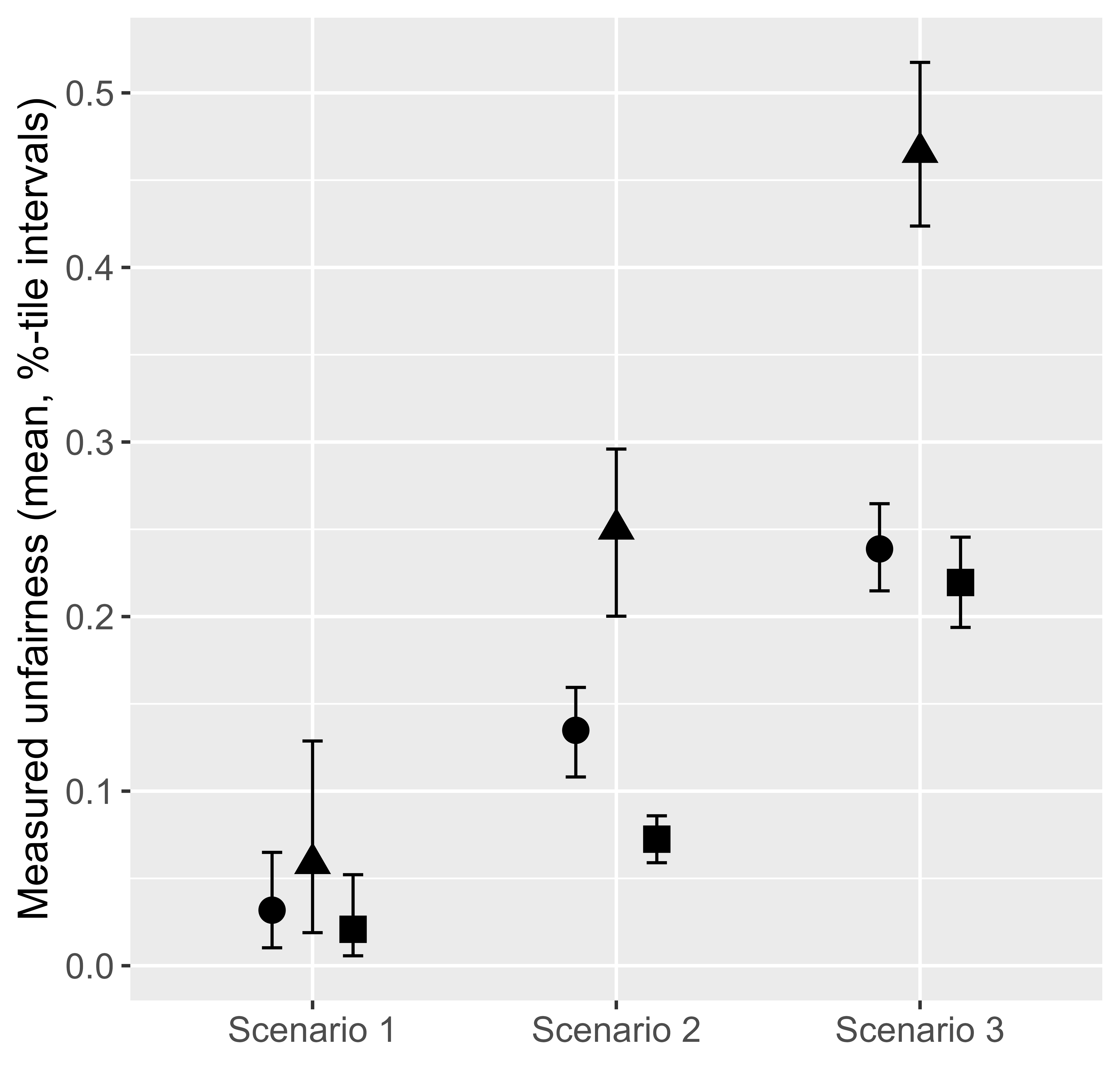}
         \caption{$\Delta_{AVG}^+$, $\Delta_{MAX}^+$, $\Delta_{VAR}^+$}
         \label{fig:sim_part1_a}
     \end{subfigure}
     \hfill
     \begin{subfigure}[b]{0.5\textwidth}
         \centering
         \includegraphics[width=\textwidth]{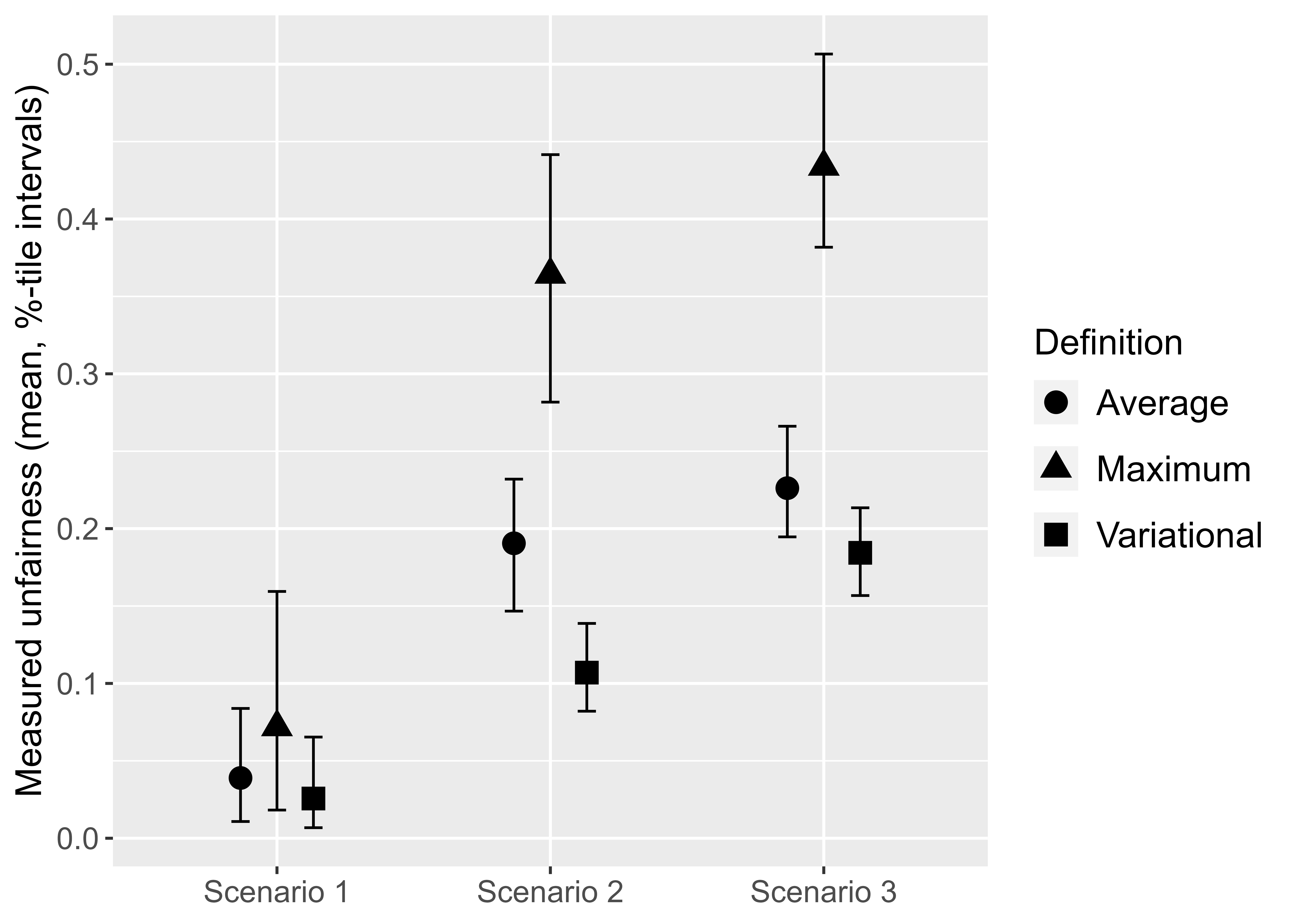}
         \caption{$\Delta_{AVG}^-$, $\Delta_{MAX}^-$, $\Delta_{VAR}^-$}
         \label{fig:sim_part1_b}
     \end{subfigure}
    \caption{Comparisons of positive (left) and negative (right) versions of new metrics. Shapes show the mean of $500$ replications of the estimation procedure, with $N_{estimation} = 9,000$ and the GLM propensity score model estimation method. Intervals show the $0.025$ and $0.975$ quantiles of the replications.}
    \label{fig:sim_part1}
\end{figure}

This section compares our new unfairness metrics in each of the three scenarios. In Scenario 1, with little unfairness, the three metrics are clustered together at a low value in both their positive (Figure \ref{fig:sim_part1_a}) and negative (Figure \ref{fig:sim_part1_b}) versions. In Scenario 2, differences emerge between $\Delta_{AVG}$ and $\Delta_{VAR}$, with the relatively lower value of $\Delta_{VAR}$ reflecting more even spacing of the error rates compared to Scenario 3. In this scenario, $\Delta_{MAX}$ is larger to reflect the large error rate differences between the minority and $M2$ and the other two groups. In Scenario 3, $\Delta_{MAX}$ is even larger because of the larger error rate differences between the majority and all other groups. However, $\Delta_{AVG}$ and $\Delta_{VAR}$ are closer together because the error rates are less evenly spaced.

These simulations use the GLM propensity score estimation method. Throughout our simulations, we found that all estimation methods proposed in Section \ref{sec:estimation} increase in precision with increasing sample size and correctly estimate the true value in Scenario 2. In Scenario 1 (low unfairness), all methods overestimate the true unfairness at low sample sizes, but all methods approach the correct value as the sample size increases. Further simulations comparing estimation methods, including under a mis-specified propensity score model, are given in the Supplementary Material.

\subsection{Inference}

\begin{figure}[h]
    \centering
    \begin{subfigure}[b]{0.38\textwidth}
         \centering
         \includegraphics[width=\textwidth]{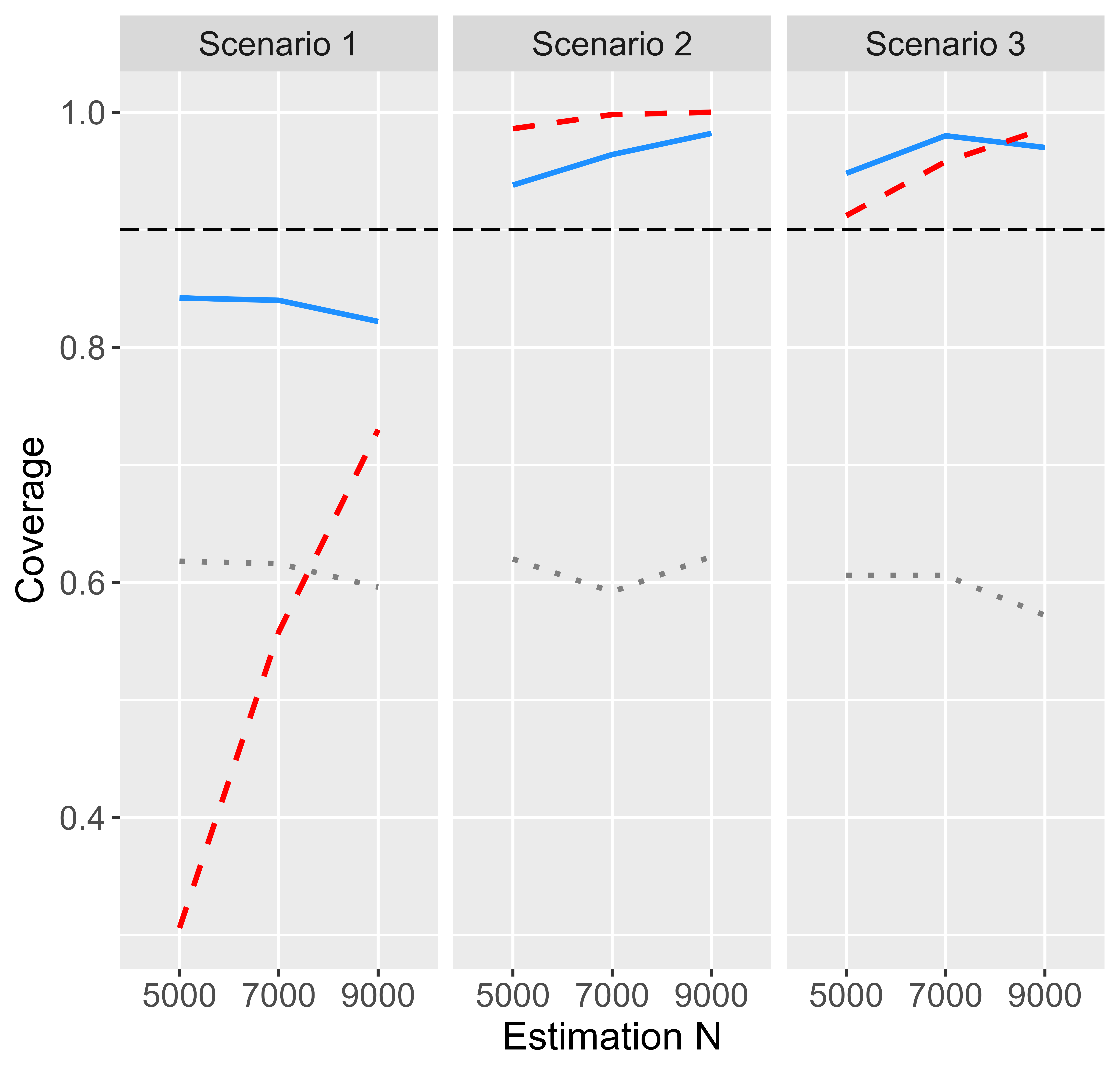}
         \caption{Coverage rates of $90\%$ confidence intervals for $\widehat{\Delta}_{AVG}^-$ using three methods.}
         \label{fig:sim_part3_a}
     \end{subfigure}
     \hfill
     \begin{subfigure}[b]{0.51\textwidth}
         \centering
         \includegraphics[width=\textwidth]{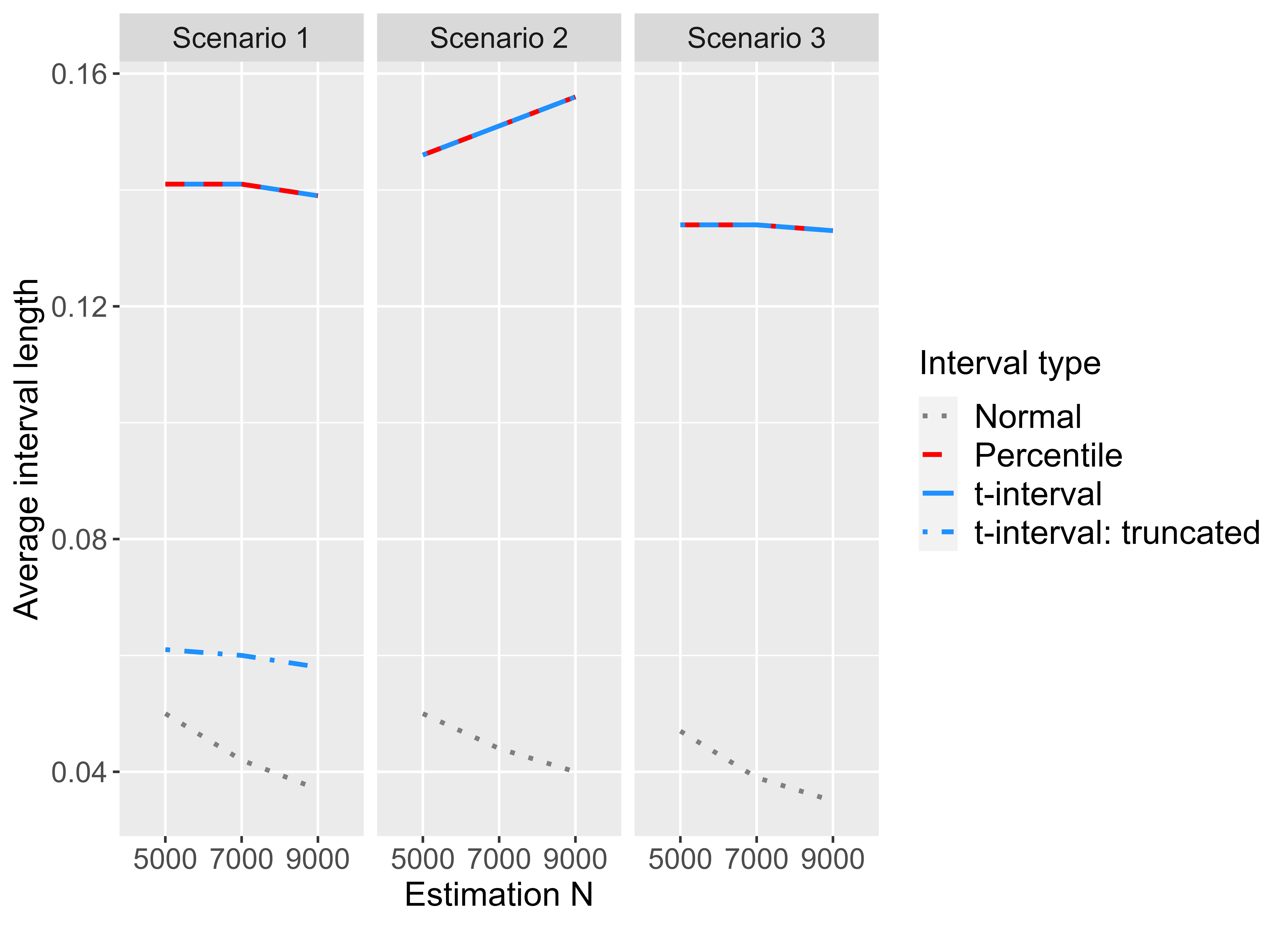}
         \caption{Average length of $90\%$ confidence intervals for $\widehat{\Delta}^-_{AVG}$ using three methods.}
         \label{fig:sim_part3_b}
     \end{subfigure}
    \caption{Lines show the performance of three methods for estimating confidence intervals, all using GLM propensity score estimation. Methods shown are normal approximation (dotted), percentile (dashed), and t-interval (solid). For each combination of scenario and sample size, coverage rates and average length of the $90\%$ confidence intervals are calculated using $1,000$ rescaled bootstrap resamples on each of $500$ estimation data sets. The horizontal dashed line in Figure \ref{fig:sim_part3_a} shows the nominal coverage rate of $90\%$. The average length of the t-intervals truncated at zero is also shown, since only the non-negative portion indicates plausible values for $\widehat{\Delta}^-_{AVG}$.}
    \label{fig:sim_part3}
\end{figure}

Figure \ref{fig:sim_part3} shows the performance of several methods for constructing confidence intervals using the rescaled bootstrap, again focusing on the negative version of $\Delta_{AVG}$. Of the three methods, only the t-interval has coverage close to the nominal rate in Scenario 1 (low unfairness) at smaller sample sizes. All methods except the normal approximation have coverage above the nominal rate in Scenarios 2 and 3 (more unfairness). Increasing sample size improves coverage for all methods and most dramatically for the percentile interval. The normal approximation gives the shortest intervals in all scenarios. The percentile and t-interval methods give intervals of the same length; however, when the t-intervals are truncated at zero in Scenario 1, their length is more comparable to that of the normal intervals. In Scenarios 2 and 3, truncation has no effect since the t-intervals do not cross zero. Increasing sample size shortens intervals in most cases, except the percentile and t-interval in Scenario 2.

\FloatBarrier
\section{Study of fairness of a COVID-19 risk model} \label{sec:application}

We applied our framework to a COVID-19 risk prediction model deployed by a major Midwestern health system during the height of the COVID-19 pandemic. This model was used to identify acute care patients at high risk of severe disease, defined as requiring intensive care admission or invasive mechanical ventilation, or resulting in death. Model scores informed decisions on whether to transfer patients to one of the system's COVID-19 cohort hospitals. The risk model was trained on 1,469 adult patients who tested positive for SARS-CoV-2 within 14 days of acute care.

We evaluated this risk model on a data set of 3,519 adult acute care patients from the same health system who tested positive for SARS-CoV-2 between 10/28/2020 and 12/31/2021. We calculated our metrics using the outcome 30-day inpatient readmission or mortality, with transfer to a cohort hospital as the treatment variable. We chose a cutoff value of $0.15$ for dichotomizing the risk score, which is approximately the $80^{th}$ percentile in our population. We examined the intersecting protected characteristics age group (54 and under, 55+) and race (white, Black or African American, other). We chose these variables since COVID-19 has been shown to unequally impact Black patients and other patients of color because of societal inequities and discrimination (e.g., \cite{aleemNewCDCData2020}), and, simultaneously, typical disease severity is understood to differ based on a patient's age. The population was predominantly white ($79\%$ white, $14\%$ Black/African American, $7\%$ other) and approximately evenly split between the age categories. We excluded patients who were missing the risk score or either protected characteristic.

For propensity score modeling, we used covariates thought to influence the assignment of treatment: comorbidities, home medications prescribed within 3 months prior to COVID positive date, number of prior emergency department (ED) visits, and labs and vitals (heart rate, oxygen saturation, respiration rate, blood pressure, temperature) collected 48 hours following ED presentation. Covariates with greater than 2/3 missing were excluded, and random forest imputation (missForest package) was performed. We considered random forest and GLM propensity score models and selected the GLM because it produced better overlap and more similar distributions between the weighted untreated and overall populations (see Supplementary Material). After propensity score modeling, we excluded 19 patients with extreme propensity scores ($>0.7$) since these patients are highly likely to receive the treatment regardless of risk score and are thus less relevant to the analysis. To maintain adequate sample size in the rescaled bootstrap resamples, we used $m=\lfloor n^{0.85} \rfloor $. All analyses were done in R (version 4.1.3, \cite{rcoreteamLanguageEnvironmentStatistical2022}).

\begin{figure}
    \centering
    \includegraphics[width=.9\textwidth]{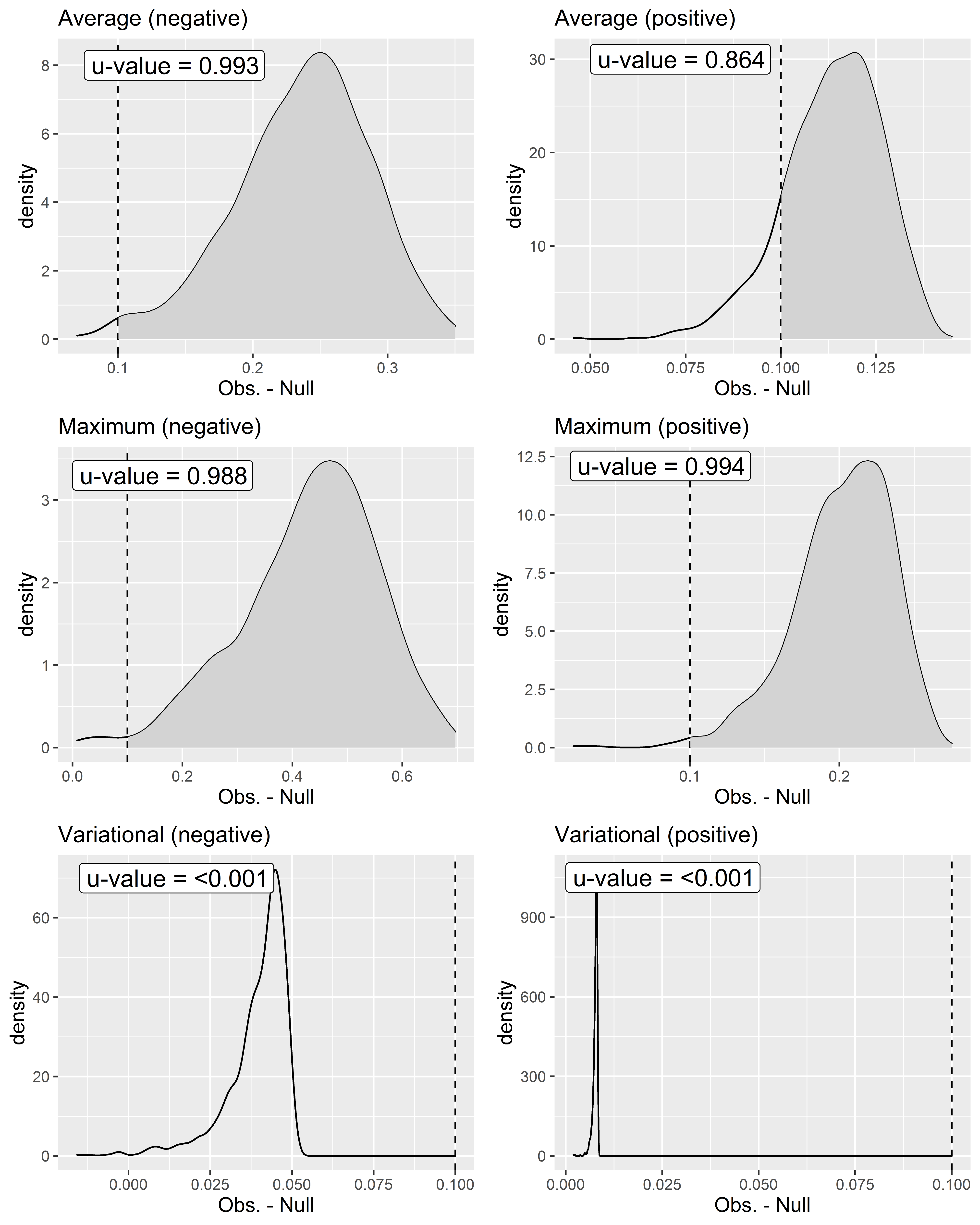}
    \caption{Application of proposed unfairness metrics to the COVID-19 risk prediction model. Density plots show observed minus null values for each metric. We used a threshold of $\delta = 0.1$ for the u-value test, giving the rounded u-values shown in the text boxes.}
    \label{fig:app_nullall}
\end{figure}

\begin{figure}[h]
    \centering
    \begin{subfigure}[b]{0.45\textwidth}
         \centering
         \includegraphics[width=\textwidth]{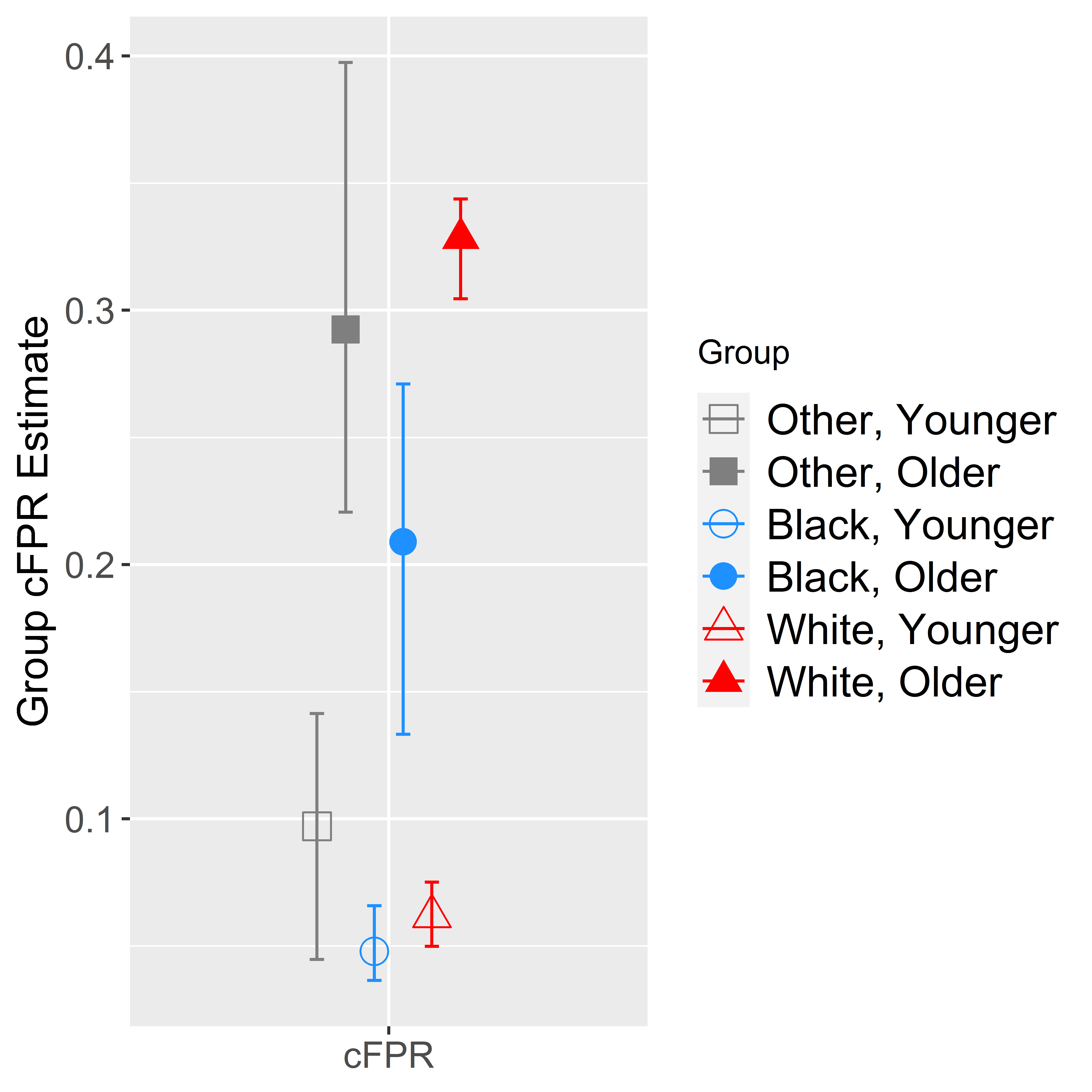}
         \caption{Counterfactual false positive rates}
         \label{fig:app_cfpr}
     \end{subfigure}
     \hfill
     \begin{subfigure}[b]{0.45\textwidth}
         \centering
         \includegraphics[width=\textwidth]{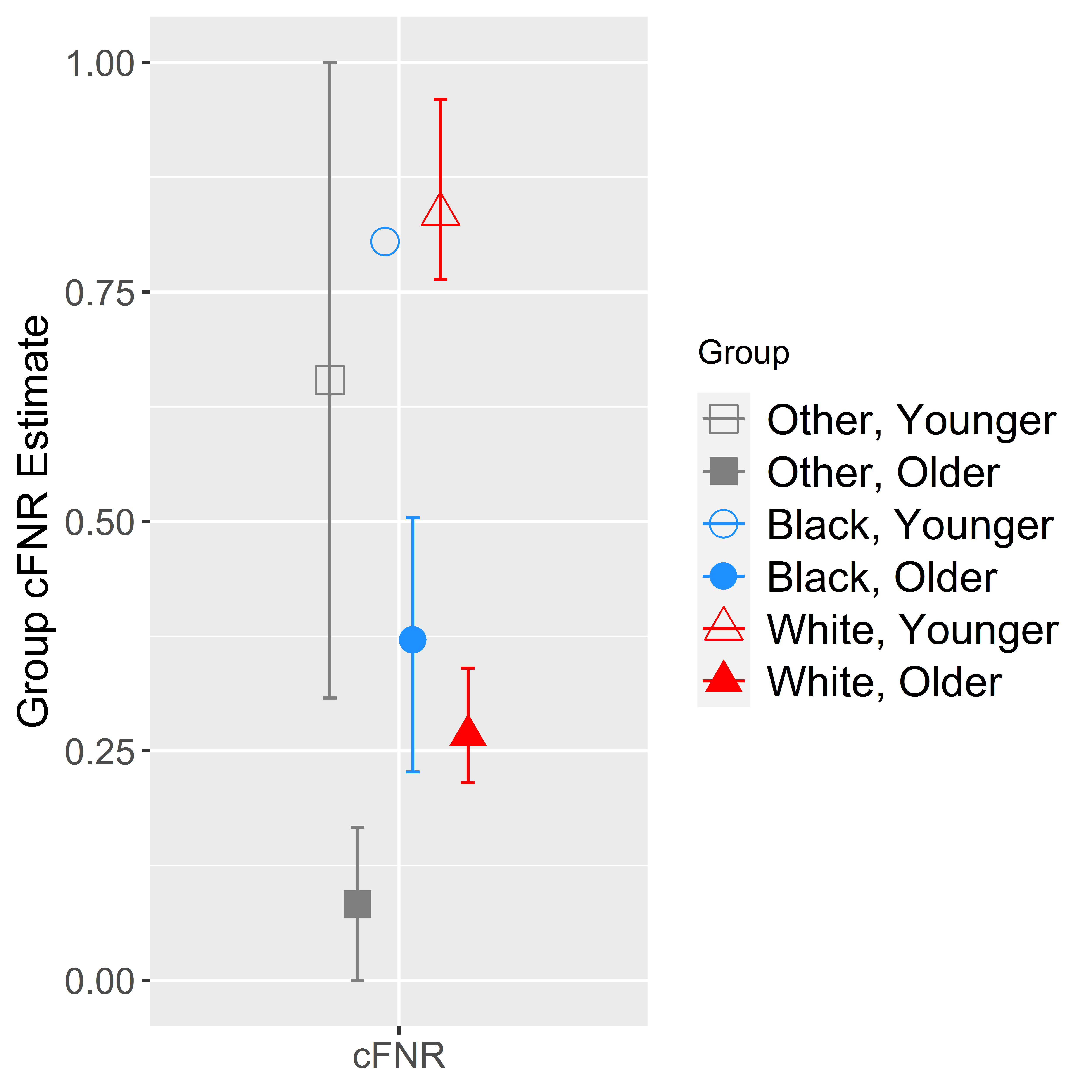}
         \caption{Counterfactual false negative rates}
         \label{fig:app_cfnr}
     \end{subfigure}
    \caption{Estimated counterfactual error rates by protected group for the COVID-19 risk prediction model. Shapes show error rate estimates. Intervals show $95\%$ t-intervals (truncated at zero) estimated using the techniques of Section \ref{sec:inference}.}
    \label{fig:app_cerrrates}
\end{figure}

Our average and maximum metrics indicated unfairness using a threshold of $\delta = 0.1$, as demonstrated by large u-values (Figure \ref{fig:app_nullall}). For our variational metric, the u-value test rejected the null hypothesis, indicating a lack of unfairness on this metric at this threshold. The estimated counterfactual error rates for each group show that most of the unfairness comes from differences in error rates for older vs. younger patients, although there are also differences by race within the age groups (Figure \ref{fig:app_cerrrates}). Error rate estimation for some subgroups was hampered by small sample size, as indicated by the large confidence intervals (e.g. other, younger group). The absence of a $cFNR$ interval for the Black, younger group is due to the fact that bootstrap resampling was done conditional on protected group, $S$, and $Y$, and there was not sufficient sample size in this group to create variation across resamples. However, the bootstrap procedure can still be used for the aggregate metrics, due to the variation in the other groups, in spite of this limitation.

\FloatBarrier
\section{Discussion} \label{sec:discussion}

In this paper we proposed tools for measuring intersectional, counterfactual unfairness in risk prediction models. We proposed three novel counterfactual unfairness metrics that accommodate multiple intersecting protected characteristics and are valid in contexts where a risk score is used to guide treatment, and we developed a full set of inference tools for our metrics. We defined the \emph{unfairness value ("u-value")} to summarize the relative extremity of unfairness compared to a hypothetical, perfectly fair model. While we applied the u-value to our intersectional, counterfactual metrics, the permutation test and user-defined threshold we propose could easily be applied to any unfairness metrics, providing a simple tool for practitioners to assess the fairness of a risk model using desired metrics.

\subsection{A wider lens for fairness}

Our intersectional fairness methods advance the conversation in clinical risk prediction by providing more nuanced, realistic measurements of inequities in model performance. However, intersectional fairness as it is typically defined and as we have defined it here, e.g. a set of technical tools for assessing and correcting differential model performance across intersecting groups, does not in itself constitute an intersectional analysis according to the original meaning of the term in Black feminist scholarship. In order to avoid misappropriating the name, "intersectional fairness" must be just one piece of an effort that spans all stages of the model development and deployment process and illuminates the real-life power relationships that cause inequity and the political work needed to dismantle them. Rather than limit focus to technical definitions of "fairness", we follow critics such as \cite{weinbergRethinkingFairnessInterdisciplinary2022} and urge practitioners to use metrics such as ours as one piece of an effort toward equitable use of algorithms that includes considerations in problem selection, representation of affected groups, implementation, and more.

\subsection{Further directions for intersectional, counterfactual fairness measurement}

Within our work on fairness measurement, several statistical issues remain for exploration. In this work we assume the availability of sufficient sample size in each intersection of the protected characteristics. When selecting characteristics, there is a trade-off between increased nuance, obtained by considering more intersecting characteristics or disaggregating categories within a characteristic, and decreased precision or inability to obtain estimates due to insufficient sample size. In particular, our weighted estimators will become unstable when all observations in a group have low propensity scores. Our estimators also require non-zero counts in all intersections of outcome and predictor values within the untreated portion of each protected group. As seen in our simulation study, these issues are not prohibitive even at relatively small overall sample and minority group sizes. However, they may arise in practice. In such cases, one solution may be to select, in a data or context-driven manner, particular combinations of protected characteristics for which to assess fairness (\cite{wangIntersectionalityMachineLearning2022a}). Others have proposed methods for such selection which, while not applied to the clinical context, provide a guide for future work (\cite{molinaBoundingApproximatingIntersectional2022}).

More broadly, while our methods can use any protected characteristics, the selection and definition of characteristics is critical to the validity of the results in context. Particularly with socially constructed characteristics like race, assignment of group labels is not straightforward, and group definitions change over time. \cite{benthallRacialCategoriesMachine2019} argue fairness work should be careful of further entrenching such inherently unequal categories. They argue for using patterns of segregation, rather than racial categories, to guide fairness interventions. Recent work on a multidimensional measure of structural racism (\cite{chantaratIntricacyStructuralRacism2021}) provides another potential alternative categorization that could be used in metrics such as ours.

\section{Software}

Software in the form of R code, a sample data set, and documentation is available online at \url{https://github.com/swastvedt/faircfint}.

\printbibliography

\newpage
\section{Tables and Figures}

\newpage

\FloatBarrier
\section{Supplementary Material}

\subsection{Proofs and derivations}

\subsubsection{Proof of Proposition \ref{prop:identification}}

We make the following standard causal inference assumptions. Let $D$ denote a binary treatment assignment, $S$ a binary risk prediction, $\bm{A}$ a protected characteristic vector, and $Y^0$ the binary potential outcome under no treatment. Let $\bm{X}$ be a vector of observed covariates. Define the propensity score function for a given protected group as $\pi = \mathbb{P}(D=1 | A, \bm{X}, S)$.

\begin{itemize}
    \item [A1.] (Consistency) The observed outcome for each subject is equal to that subject's potential outcome under the treatment actually received; i.e. $Y = DY^1 + (1-D)Y^0$.
    \item [A2.] (Positivity) $\exists \delta \in (0,1) \text{ such that } \mathbb{P}(1-\pi(\bm{A},\bm{X},S) \geq \delta) = 1$.
    \item [A3.] (Ignorability) $D$ is independent of $Y^0$ conditional on $\bm{A},\bm{X},S$.
\end{itemize}

Under assumptions A1 - A3, the following holds for functions $f(S)$ and $g(Y^0)$:

\begin{align}
    E[f(S)g(Y^0)I(\bm{A}=\bm{a})] &= E\left [ \frac{Pr(D=0 | \bm{A},\bm{X},S)}{Pr(D=0 | \bm{A},\bm{X},S)} f(S)g(Y^0)I(\bm{A}=\bm{a}) \right ]  \nonumber \\
    &= E\left[\frac{1}{1-\pi(\bm{A},\bm{X},S)} f(S) g(Y^0)I(\bm{A}=\bm{a}) E(1-D|\bm{A},\bm{X},S) \right ] \nonumber \\
    &= E\left[\frac{1}{1-\pi(\bm{A},\bm{X},S)} f(S) g(Y^0)I(\bm{A}=\bm{a}) E(1-D|Y^0, \bm{A},\bm{X},S) \right ] \label{eq:prop1_A3} \\ 
    &= E\left \{ E\left[ \frac{(1-D)f(S)g(Y^0)I(\bm{A}=\bm{a})}{1-\pi(\bm{A},\bm{X},S)}  |Y^0, \bm{A},\bm{X},S \right ] \right \} \nonumber \\
    &= E\left[ \frac{(1-D)f(S)g(Y^0)I(\bm{A}=\bm{a})}{1-\pi(\bm{A},\bm{X},S)}  \right ] \label{eq:prop1_TE} \\ 
    &= E\left[ \frac{(1-D)f(S)g(Y)I(\bm{A}=\bm{a})}{1-\pi(\bm{A},\bm{X},S)}  \right ] \label{eq:prop1_A1}
\end{align}

Line \eqref{eq:prop1_A3} follows from assumption A3 (Ignorability). Line \eqref{eq:prop1_TE} follows from the law of total expectation, and line \eqref{eq:prop1_A1} follows from assumption A1 (Consistency).

We then write each of the counterfactual error rates in terms of the functions $f(\cdot)$ and $g(\cdot)$ and apply the preceding result. 

\subsubsection{Derivation of counterfactual false positive rate}

\begin{align*}
    cFPR(S,\bm{a}) &= Pr(S=1|Y^0 = 0, \bm{A} = \bm{a}) \\
    &= \frac{Pr(S=1, Y^0 = 0, \bm{A} = \bm{a})}{Pr(Y^0 = 0, \bm{A} = \bm{a})} \\
    &= \frac{E[I(S=1)I(Y^0 = 0)I(\bm{A}=\bm{a})]}{E[I(Y^0 = 0)I(\bm{A}=\bm{a})]} \\
    &= \frac{E[S(1-Y^0)I(\bm{A}=\bm{a})]}{E[(1-Y^0)I(\bm{A}=\bm{a})]} \\
    &= \frac{E\left [ \frac{(1-D)S(1-Y)I(\bm{A} = \bm{a})}{1-\pi(\bm{A}, \bm{X}, S)} \right ]}{E \left [ \frac{(1-D)(1-Y)I(\bm{A}=\bm{a})}{1-\pi(\bm{A}, \bm{X}, S)} \right ]}
\end{align*}

where the last line follows from Proposition \ref{prop:identification} with $f(S) = S$ and $g(Y^0) = 1-Y^0$.

\subsubsection{Derivation of counterfactual false negative rate}
\begin{align*}
    cFNR(S, \bm{a}) &= Pr(S=0 | Y^0=1, \bm{A}=\bm{a}) \\
    &= \frac{Pr(S=0,Y^0=1,\bm{A}=\bm{a})}{Pr(Y^0=1,\bm{A}=\bm{a})} \\
    &= \frac{E[I(S=0)I(Y^0=1)I(\bm{A}=\bm{a})]}{E[I(Y^0=1)I(\bm{A}=\bm{a})]} \\
    &= \frac{E[(1-S)Y^0I(\bm{A}=\bm{a})]}{E[Y^0I(\bm{A}=\bm{a})]} \\
    &= \frac{E \left [ \frac{(1-D)(1-S)YI(\bm{A} = \bm{a})}{1-\pi(\bm{A}, \bm{X}, S)} \right ]}{E\left[ \frac{(1-D)YI(\bm{A} = \bm{a})}{1-\pi(\bm{A}, \bm{X}, S)} \right]}
\end{align*}

where the last line follows from Proposition \ref{prop:identification} with $f(S) = 1-S$ and $g(Y^0) = Y^0$.

\subsubsection{Derivation for regression estimators of $cFPR(S,\bm{a})$ and $cFNR(S,\bm{a})$}

\begin{align}
    cFPR(S,\bm{a}) &= \frac{Pr(S=1,Y^0=0,\bm{A}=\bm{a})}{Pr(Y^0=0, \bm{A}=\bm{a})} \nonumber \\
    &= \frac{E[Pr(S=1,Y^0=0,\bm{A}=\bm{a} | \bm{X})]}{E[Pr(Y^0=0, \bm{A}=\bm{a} | \bm{X})]} \nonumber \\
    &= \frac{E[Pr(Y^0=0|S=1,\bm{A}=\bm{a},\bm{X})Pr(S=1,\bm{A}=\bm{a}|\bm{X})]}{E[Pr(Y^0=0|\bm{A}=\bm{a},\bm{X})Pr(\bm{A}=\bm{a}|\bm{X})]} \nonumber \\
    &= \frac{E[Pr(Y=0|S=1,\bm{A}=\bm{a},\bm{X}, D=0) Pr(S=1,\bm{A}=\bm{a}|\bm{X})]}{E[Pr(Y=0|\bm{A}=\bm{a},\bm{X},D=0)Pr(\bm{A}=\bm{a}|\bm{X})]} \label{eq:regest_A1} \\
    &= \frac{E[1-E[Y|S=1,\bm{A}=\bm{a},\bm{X},D=0]E[I(\bm{A}=\bm{a})S|\bm{X}]]}{E[1-E[Y|\bm{A}=\bm{a},\bm{X},D=0]E[I(\bm{A}=\bm{a})|\bm{X}]]} \nonumber \\
    &= \frac{E[ 1-E[Y|S=1,\bm{A}=\bm{a},\bm{X},D=0] I(\bm{A}=\bm{a})S ]}{E[ 1-E[Y|\bm{A}=\bm{a},\bm{X},D=0] I(\bm{A}=\bm{a}) ]} \nonumber
\end{align}

where line \eqref{eq:regest_A1} follows from assumption A1 (Consistency).

To estimate this quantity, define $\mu_0(\bm{X}, \bm{A}, S) = E[Y|\bm{X}, \bm{A}, S, D=0]$ and $\mu_0^*(\bm{X}, \bm{A}) = E[Y|\bm{X}, \bm{A}, D=0]$. We model these quantities with generalized linear models and denote the estimated probabilities $\hat{\mu}_0$ and $\hat{\mu}_0^*$. Then the regression estimator for the counterfactual false positive rate is:

\begin{equation*}
    \widehat{cFPR(S,\bm{a})} = \frac{\sum_{i=1}^n 1-\hat{\mu}_0(\bm{X}_i, \bm{A}_i, S=1) S_i I(\bm{A}_i = \bm{a})}{\sum_{i=1}^n 1-\hat{\mu}_0^*(\bm{X}_i, \bm{A}_i) I(\bm{A}_i = \bm{a})}
\end{equation*}

The regression estimator for the counterfactual false negative rate is obtained similarly:

\begin{align*}
    cFNR(S,\bm{a}) &= \frac{Pr(S=0, Y^0=1, \bm{A}=\bm{a})}{Pr(Y^0=1,\bm{A}=\bm{a})} \\
    &= \frac{E[ Pr(S=0, Y^0=1, \bm{A}=\bm{a}) |\bm{X} ]}{E[ Pr(Y^0=1,\bm{A}=\bm{a}) |\bm{X} ]} \\
    &= \frac{E[Pr(Y=1|S=0,\bm{A}=\bm{a},\bm{X}, D=0) Pr(S=0,\bm{A}=\bm{a}|\bm{X})]}{E[Pr(Y=1|\bm{A}=\bm{a},\bm{X},D=0)Pr(\bm{A}=\bm{a}|\bm{X})]} \\
    &= \frac{E[ E[Y|S=0,\bm{A}=\bm{a},\bm{X},D=0] I(\bm{A}=\bm{a})(1-S) ]}{E[ E[Y|\bm{A}=\bm{a},\bm{X},D=0] I(\bm{A}=\bm{a}) ]}
\end{align*}

Estimated with:

\begin{equation*}
    \widehat{cFNR(S,\bm{a})} = \frac{\sum_{i=1}^n \hat{\mu}_0(\bm{X}_i, \bm{A}_i, S = 0) (1-S_i) I(\bm{A}_i = \bm{a}) }{\sum_{i=1}^n \hat{\mu}_0^*(\bm{X}_i, \bm{A}_i) I(\bm{A}_i = \bm{a})}
\end{equation*}

\subsection{Data generation for Section \ref{sec:sub-demonstration} simulations}

\begin{table}[h]
    \centering
    \begin{tabular}{@{}cccccc@{}}
        \tblhead{Group & $P(Y^0=1)$ & $P(D=1 | Y^0 = 1)$ & $P(D=1 | Y^0 = 0)$ & Int. str. & $P(\bm{A} = \bm{a})$}
        Minority & $0.4$ & $0.6$ & $0.3$ & altered & 0.04\\
        M1 & $0.4$ & $0.6$ & $0.3$ & 0.2 & 0.32 \\
        M2 & $0.4$ & $0.6$ & $0.3$ & 0.2 & 0.32 \\
        Majority & $0.2$ & $0.4$ & $0.2$ & 0.2 & 0.32
    \lastline
    \end{tabular}
    \caption{Data generating parameters fixed in Section \ref{sec:sub-demonstration} simulations. We assume that treatment never increases probability of the event, i.e. $P(Y^1 = 1 | Y^0 = 0) = 0$. Observational error rates are fixed at $0.1$ (FPR) and $0.2$ (FNR) for all groups. This simulation is based on an example in \cite{mishlerFairnessRiskAssessment2021a} demonstrating the differences between counterfactual and observational error rates for a single protected characteristic.}
    \label{tab:supp_sim_truedefs}
\end{table}

This section describes the simplified simulation used to demonstrate the need for intersectional metrics in Section \ref{sec:sub-demonstration}. In the table header, "Int. str." refers to a the parameter we use to control group counterfactual error rates. This parameter, defined in \cite{mishlerFairnessRiskAssessment2021a} as the \emph{intervention strength}, is the probability of not having the event under treatment, for protected group $A = a$, given that the event would have occurred with no treatment: $P(Y^1 = 0 | Y^0 = 1, A=a)$.

To see the connection between the intervention strength and counterfactual error rates, consider a scenario with two simplifying assumptions: a prediction model with fixed observational error rates, and the assumption that treatment never increases the chance of an adverse outcome (i.e. $P(Y^1 = 1 | Y^0 = 0) = 0$). As explained in \cite{mishlerFairnessRiskAssessment2021a}, the counterfactual error rate for a given group is then driven by the extent to which potential outcomes $Y^0$ differ from observed outcomes $Y$ in that group. Because we have assumed $Y^1 = 0$ whenever $Y^0 = 0$, the only way $Y^0$ and $Y$ can differ is when $Y^0 = 1$ and $Y = 0$, or equivalently (under our causal assumptions) when $Y^0=1$, $D=1$, and $Y^1=0$. As noted in the definition, the probability that $Y^1 = 0$ given $Y^0=1$ is the intervention strength. By altering a group's intervention strength, we therefore alter counterfactual error rates for the group. Thus different intervention strengths among protected groups can create counterfactual unfairness even when a model is fair on observational measures.

\subsection{Data generation for Section \ref{sec:sim} simulations}

To simplify notation, in this section we use $\bm{A}^*$ to denote the vector of both protected characteristics and their interaction, e.g. $\bm{A}^* = (A_1, A_2, A_1A_2)$. We follow \cite{mishlerFairnessRiskAssessment2021a} in referring to $P(D=1|Y^0=1, A_1,A_2)$ as the \emph{opportunity rate} ("OR") and $P(Y^0=1|A_1,A_2)$ as the \emph{need rate} ("NR").

\subsubsection{Protected characteristics, decision, and outcomes}
\begin{align*}
    P(A_1 = a_1, A_2 = a_2) &= \text{Mult}(1,\bm{\pi}_{A1,A2}) \\
    \bm{\pi}_{A1,A2} &= (\pi_{0,0}, \pi_{1,0}, \pi_{0,1}, \pi_{1,1}) = (0.58, 0.23, 0.13, 0.06) \\
    \bm{X} &\sim N((1,-1,2,-2)^T, 0.3^2*I_4) \\
    Y^0 &\sim Ber(\max\{\min\{p_{Y0}, 0.995\}, 0.005\});& Y^1 &\sim Ber(p_{Y1}) \\
    D &\sim Ber(\max\{\min\{p_{OR}, 0.995\}, 0.005\}) \\
    Y &= (1-D)Y^0 + D*Y^1
\end{align*}

\subsubsection{Parameters for $D$}
\begin{align*}
    \text{Training data: } p_{OR} &= \text{expit}[\text{logit}(OR_{maj}) + (X_1, X_2, \bm{A}^*)*(1,1,\bm{\beta}_{A,OR})^T] \\
    \text{Estimation data: } p_{OR} &= \text{expit}[\text{logit}(OR_{maj}) + (X_1, X_2, \bm{A}^*, S)*(1,1,\bm{\beta}_{A,OR}, \text{logit}(0.1))^T] \\
    \bm{\beta}_{A,OR} &= \begin{multlined}[t] [\text{logit}(OR_{M})-\text{logit}(OR_{maj}), \text{logit}(OR_M)-\text{logit}(OR_{maj}), \\ \text{logit}(OR_{maj})-2*\text{logit}(OR_M)+\text{logit}(OR_{min})] \end{multlined}
\end{align*}

\subsubsection{Parameters for $Y^0$ and $Y^1$}
\begin{align*}
    p_{Y0} &= \text{expit}[\text{logit}(NR_{maj}) + (\bm{X}, \bm{A}^*)*(1,1,1,1,\bm{\beta}_{A,Y0})^T] \\
    p_{Y1} &= \begin{cases} 
    Y^0 = 0 &; 0 \\
    Y^0 = 1, A_1 = 1, A_2 = 1 &; 1-z_{min} \\
    Y^0 = 1, A_1 = 1, A_2 = 0 &; 1-z_{M1} \\
    Y^0 = 1, A_1 = 0, A_2 = 1 &; 1-z_{M2} \\
    Y^0 = 1, A_1 = 0, A_2 = 0 &; 0.8 \end{cases} \\
    \bm{\beta}_{A,Y0} &= \begin{multlined}[t] [\text{\text{logit}}(NR_{M}) - \text{logit}(NR_{maj}), \text{logit}(NR_{M}) - \text{logit}(NR_{maj}), \\ \text{logit}(NR_{maj}) - 2*\text{logit}(NR_{M}) + \text{logit}(NR_{min})] \end{multlined}
\end{align*}

\subsubsection{Parameters controlling unfairness}
\begin{itemize}
    \item Scenario 1 (similar counterfactual error rates for all protected groups)
        \begin{itemize}
            \item Need rate: $NR_{maj} = 0.6$, $NR_{M} = 0.5$, $NR_{min} = 0.4$
            \item Opportunity rate: $OR_{maj} = 0.2$, $OR_{M} = 0.4$, $OR_{min} = 0.6$
            \item Intervention strength: $z_{M1} = 0.2$, $z_{M2} = 0.2$, $z_{min} = 0.6$
            \item Predictors for random forest risk prediction model: $\bm{X}$ 
        \end{itemize} 
    \item Scenario 2 (Unfairness involving multiple groups)
        \begin{itemize}
            \item Need rate and Opportunity rate: same as Scenario 1
            \item Intervention strength: $z_{M1} = 0.3$, $z_{M2} = 0.4$, $z_{min} = 0.5$
            \item Predictors for random forest risk prediction model: $A_1$, $A_2$, $\bm{X}$ 
        \end{itemize}
    \item Scenario 3 (Unfairness involving one group)
        \begin{itemize}
            \item Need rate: $NR_{maj} = 0.8$, $NR_{M} = 0.4$, $NR_{min} = 0.4$
            \item Opportunity rate: $OR_{maj} = 0.4$, $OR_{M} = 0.6$, $OR_{min} = 0.6$
            \item Intervention strength: $z_{M1} = 0.2$, $z_{M2} = 0.2$, $z_{min} = 0.2$ 
            \item Predictors for random forest risk prediction model: $A_1$, $A_2$, $\bm{X}$
        \end{itemize}
\end{itemize}

\FloatBarrier
\subsection{Additional simulations}

\subsubsection{Comparison of estimation methods}

\begin{figure}[h]
     \includegraphics[width=0.9\textwidth]{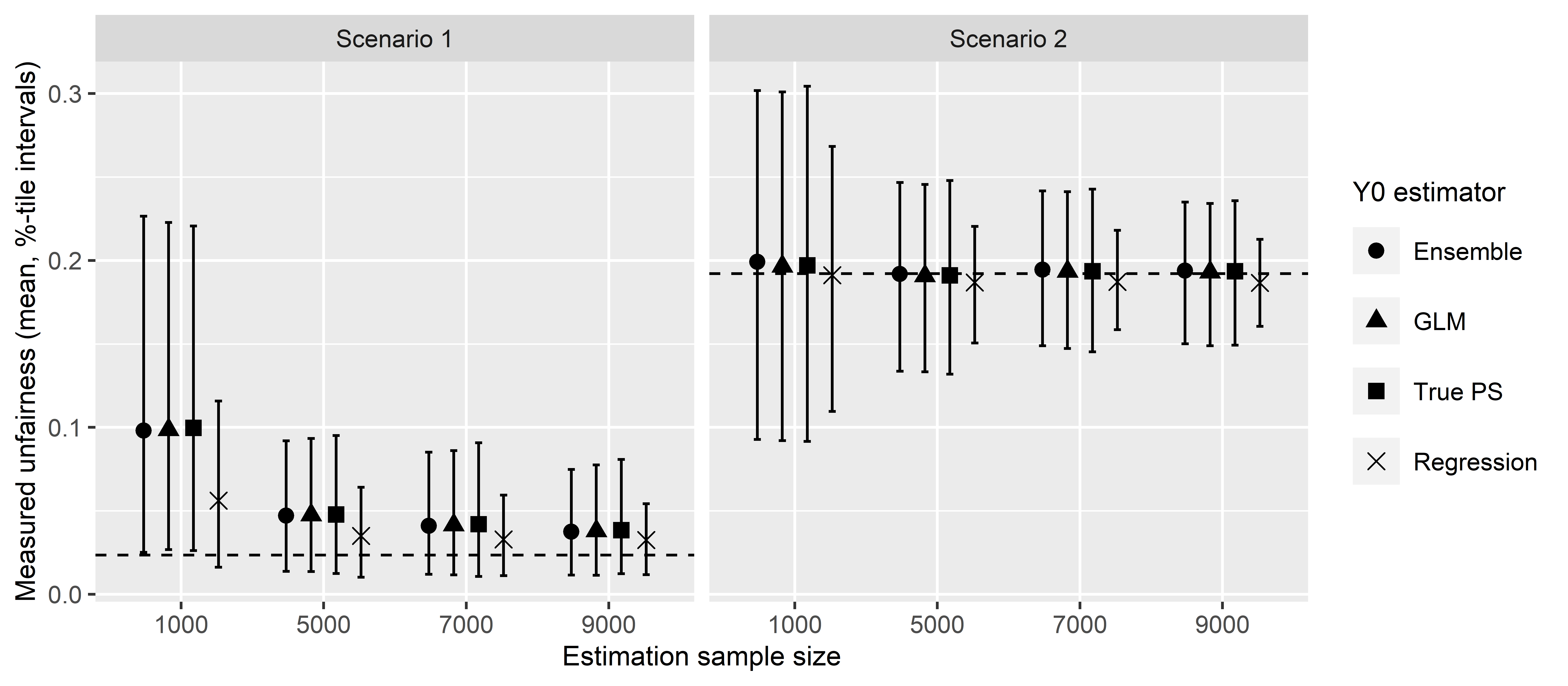}
      \caption{Performance of estimation methods for the negative version of $\Delta_{AVG}$ under Scenario 1 (low unfairness) and Scenario 2 (more unfairness). Shapes show the mean of $500$ replications of the estimation procedure, and error bars show the $0.025$ and $0.975$ quantiles. The true values, as obtained from the validation data sets for each scenario, are shown with the horizontal dotted lines. Estimation methods considered are the regression estimator ("Regression") and weighted estimators using ensemble and GLM propensity score models. A weighted estimator using the true propensity score model ("True PS") is included for comparison.}
     \label{fig:sim_part2}
\end{figure}

Figure \ref{fig:sim_part2} demonstrates the performance of the estimation methods proposed in Section \ref{sec:estimation}. We focus on the negative version of $\Delta_{AVG}$ and show scenarios 1 and 2 to demonstrate performance with both near-zero and higher unfairness. As the estimation sample size increases, the precision of the estimates increases for all methods and scenarios, as shown by the decreasing lengths of the error bars. In Scenario 1 (low unfairness), all methods overestimate the true unfairness at low sample sizes, but all methods approach the correct value as the sample size increases. For this data, in which the true data generating mechanism is known and the regression estimator can be correctly specified, it performs slightly better than the weighted estimators at lower sample sizes.

In Scenario 2 (more unfairness), all methods correctly approximate the true value at all sample sizes. Again, with the regression model correctly specified, it returns slightly shorter intervals than the weighted methods.

\subsubsection{Mis-specified propensity score model}

We explore the effect of a mis-specified propensity score model on our estimators using the simulation set-up of \cite{kangDemystifyingDoubleRobustness2007a}, in which the propensity score model is constructed using transformed covariates. We alter the data generation of Section \ref{sec:sim} such that $p_{Y0} = expit[(\pmb{X}, \pmb{A}^*)*(1,-0.75,-1,0.25, \beta_{A,Y0})^T]$ with $\beta_{A,Y0} = (0.5,-0.5,-0.1)$ and $p_{OR} = expit[logit(0.2) + (\pmb{X},\pmb{A}^*,S)*(1,0.75,-0.25,0.5,\beta_{A,OR}, 0.5)^T]$ for the estimation data with $\beta_{A,OR} = (0.5,0.5,0.2)$. For the training data, $p_{OR}$ is as in the estimation data but removing $S$. 

We model the propensity score using a GLM and the transformed covariates $\pmb{X}^*$:

\begin{itemize}
    \item $X_1^* = exp(X_1/2)$
    \item $X_2^* = X_2/(1+exp(X_1))+10$
    \item $X_3^* = (X_1X_3/25 + 0.6)^3$
    \item $X_4^* = (X_2+X_4+20)^2$
\end{itemize}

\begin{figure}[h]
    \centering
   \begin{subfigure}[b]{0.8\textwidth}
         \centering
         \includegraphics[width=\textwidth]{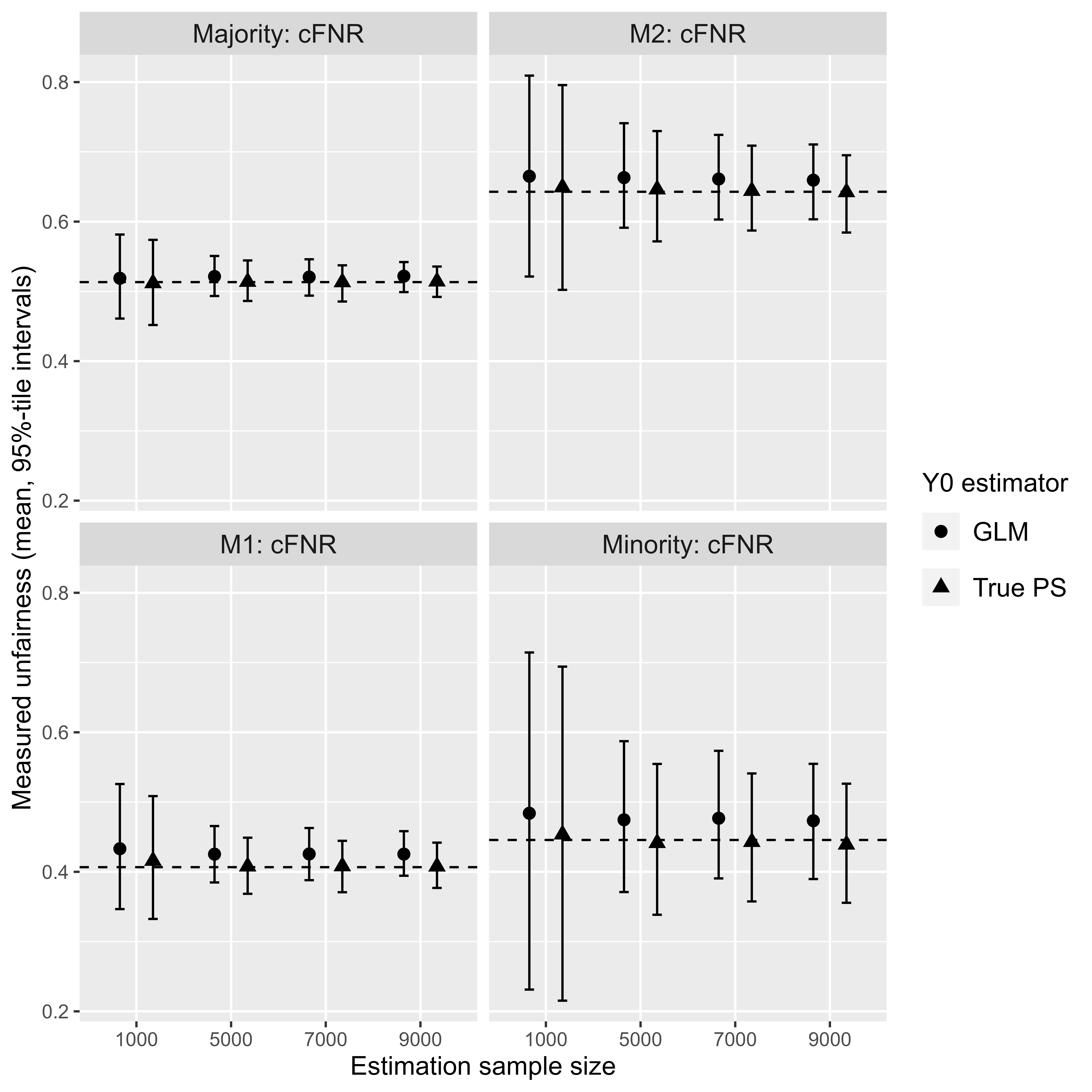}
         \caption{Estimated group counterfactual false negative rates}
         \label{fig:sim_misspec_a}
     \end{subfigure}
     \hfill
     \begin{subfigure}[b]{0.8\textwidth}
         \centering
         \includegraphics[width=\textwidth]{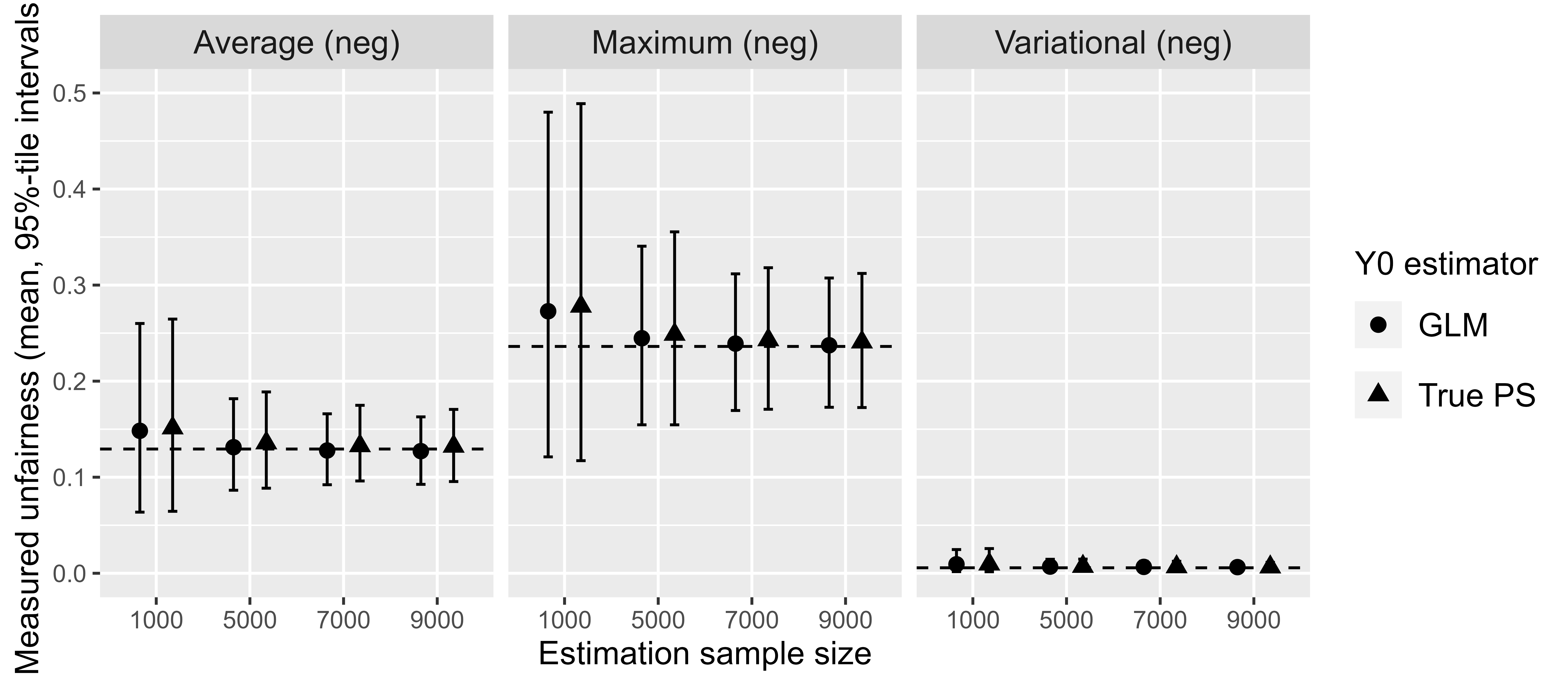}
         \caption{Estimated summary metrics}
         \label{fig:sim_misspec_b}
     \end{subfigure}
    \caption{Performance of estimation methods under a mis-specified propensity score model. Dotted horizontal lines show the true values, as obtained from the validation data set. Shapes show the mean of $500$ replications of the estimation procedure, and error bars show the $0.025$ and $0.975$ quantiles. }
    \label{fig:sim_misspec}
\end{figure}

Figure \ref{fig:sim_misspec} demonstrates that a mis-specified propensity score model cases bias in estimation of the group counterfactual error rates. Although in this case the bias cancels out when the group error rates are aggregated into our summary metrics, this will not be the case in general. As we discuss in Section \ref{sec:estimation-nuissance}, as with all IPW estimators, our metrics rely on correct specification of the propensity score model.

\FloatBarrier
\subsection{Doubly robust estimation of summary metrics}

As mentioned in the main text, we prefer IPW estimators of the counterfactual error rates and our summary metrics because of such estimators' adaptability and suitability to applications in which modeling the propensity score is much more feasible than modeling the outcome (Section \ref{sec:estimation-nuissance}). However, in cases where both models are obtainable, practitioners may wish to use doubly robust estimators. To obtain doubly robust estimators of $cFPR(S,\pmb{a})$ and $cFNR(S,\pmb{a})$, we use the following result which is true under assumptions A1 - A3 from Section \ref{prop:identification}.

\begin{proposition}\label{prop:supp_1}
Given arbitrary functions $f(S)$ and $g(Y^0)$ with finite mean, the following holds:

\begin{equation*}
\begin{multlined}[t]E[f(S)g(Y^0)I(\pmb{A} = \pmb{a})] = E [f(S)I(\pmb{A}=\pmb{a}) \{ m(g(Y),f(S),\pmb{a},\pmb{X}) + \\ \frac{(1-D)}{1-\pi(\pmb{A},\pmb{X},S)} \{Y - m(g(Y),f(S),\pmb{a},\pmb{X}) \} \} ]\end{multlined}
\end{equation*}
where $m(g(Y),f(S),\pmb{a},\pmb{X}) = E[g(Y)|D=0,f(S),\pmb{A}=\pmb{a},\pmb{X}]$
\end{proposition}

\textbf{Proof:} Let $p_0 = E[f(S)g(Y^0)I(\pmb{A}=\pmb{a})|\pmb{X}]$. Then the following holds:

\begin{align}
    E[f(S)g(Y^0)I(\pmb{A}=\pmb{a})] &= E[E[f(S)g(Y^0)I(\pmb{A}=\pmb{a})|\pmb{X}]] + E[f(S)g(Y^0)I(\pmb{A}=\pmb{a})] - E[f(S)g(Y^0)I(\pmb{A}=\pmb{a})] \nonumber \\
    &= E[p_0] + E\left [ \frac{Pr(D=0|\pmb{A},\pmb{X},S)}{Pr(D=0|\pmb{A},\pmb{X},S)} f(S)g(Y^0) I(\pmb{A}=\pmb{a}) \right ] - E[p_0] \nonumber \\
    &= E[p_0] + E\left [ \frac{E(1-D|\pmb{A},\pmb{X},S)}{1-\pi(\pmb{A},\pmb{X},S)} f(S)g(Y^0) I(\pmb{A}=\pmb{a}) \right ] - E \left[ \frac{E(1-D|\pmb{A},\pmb{X},S)}{1-\pi(\pmb{A},\pmb{X},S)} p_0 \right] \nonumber \\
    &= \begin{multlined}[t]
         E[p_0] + \left [ E \left (\frac{1-D}{1-\pi(\pmb{A},\pmb{X},S)} f(S)g(Y^0) I(\pmb{A}=\pmb{a}) | Y^0, \pmb{A},\pmb{X},S \right ) \right] -  \\ E\left [ E \left ( \frac{1-D}{1-\pi(\pmb{A},\pmb{X},S)}p_0|\pmb{A},\pmb{X},S \right )\right ] 
         \end{multlined} \nonumber \\
    &= E[p_0] + E \left [ \frac{1-D}{1-\pi(\pmb{A},\pmb{X},S)} f(S)g(Y^0) I(\pmb{A}=\pmb{a}) \right ] - E\left[ \frac{1-D}{1-\pi(\pmb{A},\pmb{X},S)} p_0 \right] \nonumber \\
    &= E \left[ p_0 + \frac{1-D}{1-\pi(\pmb{A},\pmb{X},S)} \{f(S)g(Y) I(\pmb{A}=\pmb{a}) - p_0\} \right ] \label{eq:supp_dr1}
\end{align}

Then re-write $p_0$ as:

\begin{align*}
    p_0 &= E[f(S)g(Y^0)I(\pmb{A}=\pmb{a})|\pmb{X}] \\
    &= E[E(g(Y^0)|f(S),I(\pmb{A}=\pmb{a}),\pmb{X})E(f(S)I(\pmb{A}=\pmb{a})|\pmb{X})] \\
    &= E[E(g(Y)|D=0,f(S),\pmb{A}=\pmb{a},\pmb{X})E(f(S)I(\pmb{A}=\pmb{a})|\pmb{X})]] \\
    &= E[E(g(Y)|D=0,f(S),\pmb{A}=\pmb{a},\pmb{X})f(S)I(\pmb{A}=\pmb{a})] \\
    & E[f(S)I(\pmb{A}=\pmb{a}) m(g(Y),f(S),\pmb{a},\pmb{X})]
\end{align*}

Finally, simplify equation \ref{eq:supp_dr1} as:

\begin{equation*}
    E \left[ f(S)I(\pmb{A}=\pmb{a}) \left \{ m(g(Y),f(S),\pmb{a},\pmb{X}) + \frac{1-D}{1-\pi(\pmb{A},\pmb{X},S)} \{Y - m(g(Y),f(S),\pmb{a},\pmb{X})\}  \right \}\right ]
\end{equation*}

In the case of $cFNR(S,\pmb{a})$ and $cFPR(S,\pmb{a})$, let $m(Y,S,\pmb{a},\pmb{X}) = P(Y=1|S,\pmb{A}=\pmb{a},\pmb{X})$ and let $m(Y,\pmb{a},\pmb{X}) = P(Y=1|\pmb{A}=\pmb{a},\pmb{X})$. Then we have the following.

\begin{align*}
    cFNR(S,\pmb{a}) &= \dfrac{E \left[ (1-S)I(\pmb{A}=\pmb{a}) \left \{ m(Y,S,\pmb{a},\pmb{X}) + \frac{1-D}{1-\pi(\pmb{A},\pmb{X},S)} \{Y - m(Y,S,\pmb{a},\pmb{X})\}  \right \}\right ]}{E\left[I(\pmb{A}=\pmb{a}) \left \{ m(Y,\pmb{a},\pmb{X}) +\frac{1-D}{1-\pi(\pmb{A},\pmb{X},S)}\{Y-m(Y,\pmb{a},\pmb{X})\} \right \} \right]]} \\
    cFPR(S,\pmb{a}) &= \dfrac{E \left[ S I(\pmb{A}=\pmb{a}) \left \{1-m(Y,S,\pmb{a},\pmb{X}) + \frac{1-D}{1-\pi(\pmb{A},\pmb{X},S)} \{m(Y,S,\pmb{a},\pmb{X}) - Y \}  \right \}\right ]}{E \left[I(\pmb{A}=\pmb{a}) \left \{ (1-m(Y,\pmb{a},\pmb{X})) + \frac{1-D}{1-\pi(\pmb{A},\pmb{X},S)} \{m(Y,\pmb{a},\pmb{X})-Y\}  \right \}\right ]}
\end{align*}

Following \cite{robinsCommentPerformanceDoubleRobust2007}, we use parametric models for $m$, and $\pi$ and denote the versions of $m$ estimated using weighted least squares as $\hat{m}(S,\pmb{a},\pmb{x})_{WLS}$ and $\hat{m}(\pmb{a},\pmb{x})_{WLS}$. We estimate the above quantities using these models and sample averages for the population expectations.

These estimators are doubly robust since they are of the form $\hat{\mu}_{DR}(\hat{\pi},\hat{m})$ given in \cite{robinsCommentPerformanceDoubleRobust2007}, with the addition of indicator variables $S$ (or $1-S$) and $I((\pmb{A} = \pmb{a})$ in each sample average. Further, \cite{robinsCommentPerformanceDoubleRobust2007} note that when weighted least squares is used to estimate $\hat{m}$ and the model for $m$ has an intercept, the resulting estimator satisfies $\mathbb{P}_n \left [\frac{1-D}{1-\hat{\pi}} \{Y-\hat{m}_{WLS}\} \right ] = 0$. Thus the second component of the general form in Proposition \ref{prop:supp_1} is zero when estimated in this way.

This gives the following doubly robust estimators for the counterfactual error rates, which can be substituted in error rate difference expressions and our summary metrics in Section \ref{sec:estimation} of the main text. As \cite{robinsCommentPerformanceDoubleRobust2007} note, estimators of this form are bounded in $[0,1]$, the natural range for error rates.

\begin{align*}
    \widehat{cFPR}(S,\pmb{a})_{DR} &= \frac{\sum_{i=1}^n S_iI\{\pmb{A}_i=\pmb{a}\} (1-\hat{m}_{WLS}(s_i,\pmb{a},\pmb{x}_i))}{\sum_{i=1}^n I\{\pmb{A}_i=\pmb{a}\} (1-\hat{m}_{WLS}(\pmb{a},\pmb{x}_i)) } \\
    \widehat{cFNR}(S,\pmb{a})_{DR} &= \frac{\sum_{i=1}^n (1-S_i)I\{\pmb{A}_i=\pmb{a}\} \hat{m}_{WLS}(s_i,\pmb{a},\pmb{X}_i)}{\sum_{i=1}^n I\{\pmb{A}_i=\pmb{a}\} \hat{m}_{WLS}(\pmb{a},\pmb{X}_i) }
\end{align*}

\FloatBarrier
\subsection{Overlap plot for COVID-19 application}

\begin{figure}[h]
    \centering
    \includegraphics[width=.6\textwidth]{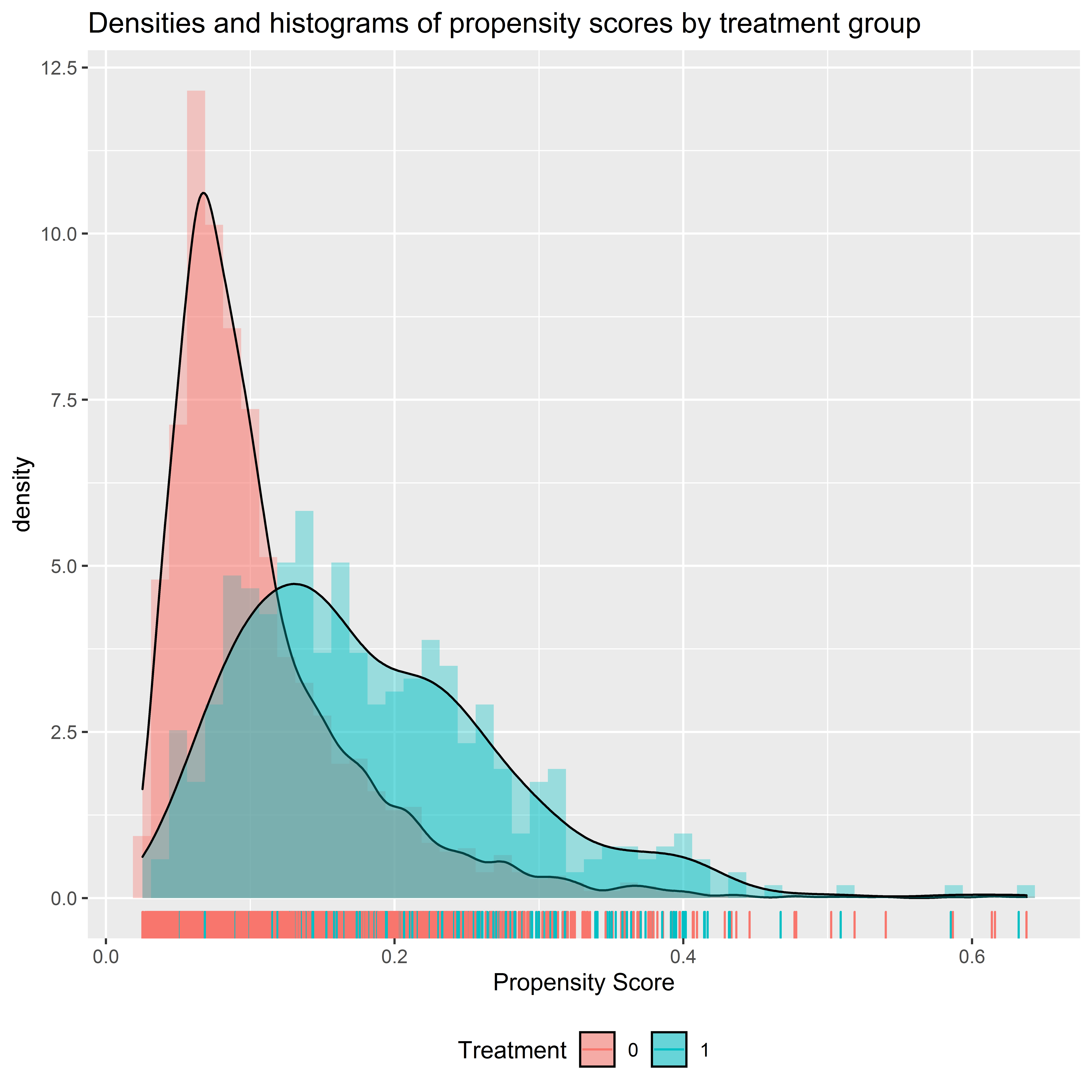}
    \caption{Propensity score overlap for two treatment groups in the COVID-19 risk prediction model application (Treatment $1$: transfer to cohort hospital). Density plots show the distribution of propensity scores for the GLM model by treatment group after excluding 19 observations with propensity scores $> 0.7$. (Plot produced with \emph{check.overlap}, package \emph{personalized})}
    \label{fig:supp_overlap}
\end{figure}
\end{document}